\documentclass[epj]{svjour}
\usepackage[dvips]{graphicx}
\usepackage{subeqnarray}
\usepackage{amsmath,amsfonts}
\graphicspath{{ps/}{eps/}{epsi/}{./}{../ps/}{../eps/}
		{/graal/throng/article_KL2/figures/}}

\begin{document}

\title{Measurement of beam-recoil observables $O_x$ , $O_z$ and target asymmetry for the reaction $\gamma p \rightarrow K^+
\Lambda$}

\author{
A.~Lleres\inst{1},
O.~Bartalini\inst{2,10},
V.~Bellini\inst{13,6},
J.P.~Bocquet\inst{1},
P.~Calvat\inst{1},
M.~Capogni\inst{2,10,4},
L.~Casano\inst{10},
M.~Castoldi\inst{8},
A.~D'Angelo\inst{2,10},
J.-P.~Didelez\inst{16},
R.~Di~Salvo\inst{10},
A.~Fantini\inst{2,10},
D.~Franco\inst{2,10},
C.~Gaulard\inst{5,14},
G.~Gervino\inst{3,11},
F.~Ghio\inst{9,12},
B.~Girolami\inst{9,12},
A.~Giusa\inst{13,7},
M.~Guidal\inst{16},
E.~Hourany\inst{16},
R.~Kunne\inst{16},
V.~Kuznetsov\inst{15,18},
A.~Lapik\inst{15},
P.~Levi~Sandri\inst{5},
F.~Mammoliti\inst{13,7},
G.~Mandaglio\inst{7,17},
D.~Moricciani\inst{10},
A.N.~Mushkarenkov\inst{15},
V.~Nedorezov\inst{15},
L.~Nicoletti\inst{2,10,1},
C.~Perrin\inst{1},
C.~Randieri\inst{13,6},
D.~Rebreyend\inst{1},
F.~Renard\inst{1},
N.~Rudnev\inst{15},
T. Russew\inst{1},
G. Russo\inst{13,7},
C.~Schaerf\inst{2,10},
M.-L.~Sperduto\inst{13,7},
M.-C.~Sutera\inst{7},
A.~Turinge\inst{15}
(The GRAAL collaboration)
}
\offprints{lleres@lpsc.in2p3.fr}   
\institute{LPSC, Universit\'e Joseph Fourier Grenoble 1, CNRS/IN2P3, Institut National Polytechnique de Grenoble, 53 avenue des Martyrs, 38026 Grenoble, France
\and 
Dipartimento di Fisica, Universit\`a di Roma "Tor Vergata", via della Ricerca Scientifica 1, I-00133 Roma, Italy
\and
Dipartimento di Fisica Sperimentale, Universit\`a di Torino, via P. Giuria, I-00125 Torino, Italy
\and
Present affiliation: ENEA - C.R. Casaccia, via Anguillarese 301, I-00060 Roma, Italy
\and
INFN - Laboratori Nazionali di Frascati, via E. Fermi 40, I-00044 Frascati, Italy
\and
INFN - Laboratori Nazionali del Sud, via Santa Sofia 44, I-95123 Catania, Italy
\and
INFN - Sezione di Catania, via Santa Sofia 64, I-95123 Catania, Italy
\and
INFN - Sezione di Genova, via Dodecanneso 33, I-16146 Genova, Italy
\and
INFN - Sezione di Roma, piazzale Aldo Moro 2, I-00185 Roma, Italy
\and
INFN - Sezione di Roma Tor Vergata, via della Ricerca Scientifica 1, I-00133 Roma, Italy
\and
INFN - Sezione di Torino, I-10125 Torino, Italy
\and
Istituto Superiore di Sanit\`a, viale Regina Elena 299, I-00161 Roma, Italy
\and
Dipartimento di Fisica ed Astronomia, Universit\`a di Catania, via Santa Sofia 64, I-95123 Catania, Italy
\and 
Present affiliation: CSNSM, Universit\'e Paris-Sud 11, CNRS/IN2P3, 91405 Orsay, France
\and
Institute for Nuclear Research, 117312 Moscow, Russia
\and
IPNO, Universit\'e Paris-Sud 11, CNRS/IN2P3, 15 rue Georges Cl\'emenceau, 91406 Orsay, France
\and
Dipartimento di Fisica, Universit\`a di Messina, salita Sperone, I-98166 Messina, Italy
\and
Kyungpook National University, 702-701, Daegu, Republic of Korea}

\date{Received: date / Revised version: date}
\abstract{The double polarization (beam-recoil) observables $O_x$ and $O_z$ have been measured for the reaction
$\gamma p \rightarrow K^+\Lambda$ from threshold production to $E_\gamma \sim$ 1500~MeV.
The data were obtained with the linearly polarized beam of the GRAAL facility.
Values for the target asymmetry $T$ could also be extracted despite the use of an unpolarized target.
Analyses of our results by two isobar models tend to confirm the necessity to include
new or poorly known resonances in the 1900 MeV mass region.} 
\maketitle
\section{Introduction}
\label{intro}

A detailed and precise knowledge of the nucleon spectroscopy is undoubtedly one of the cornerstones for our 
understanding of the strong interaction in the non-perturbative regime. Today's privileged way to get information 
on the excited states of the nucleon is light meson photo- and electroproduction. The corresponding database 
has considerably expanded over the last years thanks to a combined effort of a few dedicated facilities worldwide. 
Not only did the recent experiments brought a quantitative improvement by measuring cross sections with 
unprecedented precision for a large number of channels but they also allowed a qualitative leap by providing 
for the first time high quality data on polarization observables. It is well known -- and now 
well established -- that these variables, being interference terms of various multipoles, bring unique 
and crucial constraints for partial wave analysis, hence facilitating the identification of resonant 
contributions and making parameter extraction more reliable.

From this perspective, $K^+\Lambda$ photoproduction offers unique opportunities. Because the 
$\Lambda$ is a self-analyzing particle, several polarization observables can be "easily" measured 
via the analysis of its decay products. As a consequence, this reaction already possesses
the richest database with results on the differential cross section \cite{bra06}-\cite{sum06}, 
two single polarization observables ($\Sigma$ and $P$) \cite{gla04}-\cite{zeg03} and two double polarization 
observables ($C_x$ and $C_z$) recently measured by the CLAS collaboration \cite{bra07}. 
On the partial wave analysis side, the situation 
is particularly encouraging with most models concluding to the necessity of incorporating 
new or poorly known resonances to reproduce the full set of data. Some discrepancies do 
remain nonetheless either on the number of used resonances or on their identification. 
To lift the remaining ambiguities, new polarization obervables are needed calling for new experiments.

In the present work, we report on first measurements of the beam-recoil observables $O_x$ and $O_z$
for the reaction $\gamma p \rightarrow K^+\Lambda$ over large energy (from threshold to 1500 MeV)
and angular ($\theta_{cm} = 30-140^0$) ranges. The target asymmetry $T$, indirectly extracted from the data, 
is also presented.
 
\section{Experimental set-up}
\label{setup}

The experiment was carried-out with the GRAAL facility (see \cite{bar05} 
for a detailed description), installed at the
European Synchrotron Radiation Facility (ESRF) in Grenoble (France). The
tagged and linearly polarized $\gamma$-ray beam is produced by Compton scattering of 
laser photons off the 6.03~GeV electrons circulating in the storage ring.

In the present experiment, we have used a set of UV lines at 333, 351 and 
364~nm produced by an Ar laser, giving 1.40, 1.47 and 1.53~GeV $\gamma$-ray 
maximum energies, respectively. Some data were also taken with the green line 
at 514~nm (maximum energy of 1.1 GeV).

The photon energy is provided by an internal tagging system. The position of the
scattered electron is measured by a silicon microstrip detector (128 strips with 
a pitch of 300~$\mu$m  and 1000~$\mu$m thick). The measured energy resolution of 
16~MeV is dominated by the energy dispersion of the electron beam
(14 MeV - all resolutions are given as FWHM). 
The energy calibration is extracted run by run from the fit of the Compton edge position with a
precision of $\sim$10$\mu m$, equivalent to
$\Delta E_\gamma/E_\gamma \simeq 2 \times 10^{-4}$ (0.3~MeV at 1.5~GeV).
A set of plastic scintillators used for time measurements is placed 
behind the microstrip detector. Thanks to a specially designed electronic module which
synchronizes the detector signal with the RF of the machine, the
resulting time resolution is $\approx$100~ps.
The coincidence between detector signal and RF is used as a start for all Time-of-Flight (ToF) 
measurements and is part of the trigger of the experiment. 

The energy dependence of the $\gamma$-ray beam polarization was determined 
from the Klein-Nishina formula taking into account the laser and electron beam emittances.
The UV beam polarization is close to 100\% at the maximum energy and decreases smoothly with
energy to around 60\% at the $K\Lambda$ 
threshold (911~MeV). Based on detailed studies \cite{bar05}, it was found that
the only significant source of error for the
$\gamma$-ray polarization $P_\gamma$ comes from the laser beam
polarization ($\delta P_\gamma / P_\gamma$=2\%).

A thin monitor is used to measure the beam flux (typically 10$^6$ $\gamma$/s). The monitor 
efficiency (2.68$\pm$0.03\%) was estimated by comparison with the response at low rate of
a lead/scintillating fiber calorimeter. 

The target cell consists of an aluminum hollow cylinder of 4~cm in diameter closed by thin
mylar windows (100~$\mu$m) at both ends. Two different target lengths (6
and 12~cm) were used for the present experiment. The target was filled by 
liquid hydrogen at 18~K ($\rho \approx 7 \ 10^{-2}$~g/cm$^3$).

\begin{figure}
\begin{center}
\includegraphics[width=0.9\linewidth]{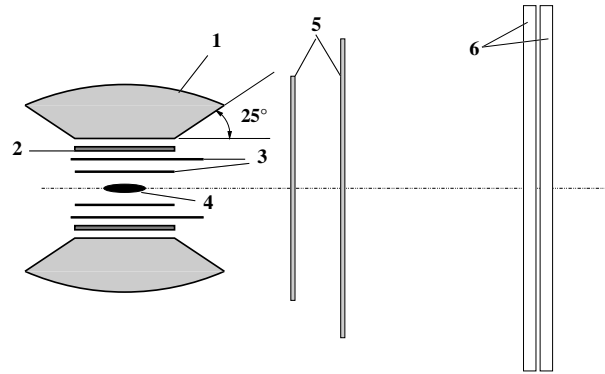} 
\end{center}
\caption{Schematic view of the LA$\gamma$RANGE detector: BGO calorimeter (1), plastic scintillator barrel (2),
cylindrical MWPCs (3), target (4), plane MWPCs (5), double plastic scintillator hodoscope (6)
(the drawing is not to scale).}
\label{sch}
\end{figure}

The 4$\pi$ LA$\gamma$RANGE detector of the GRAAL set-up allows to
detect both neutral and charged particles (fig. \ref {sch}). The
apparatus is composed of two main parts: a central one (25$^0\leq \theta \leq
155^0$) and a forward one ($\theta \ \leq \ 25^0$).

The charged particle tracks are measured by a set of  
MultiWire Proportional Chambers (MWPC) (see \cite{lle07} 
for a detailed description). 
To cover forward angles, two plane chambers,
each composed of two planes of wires, are used.
The detection efficiency of a track is about 95\% and 
the average polar and azimuthal resolutions are 1.5$^0$ and 2$^0$, respectively.
The central region is covered by two coaxial cylindrical chambers.
Single track efficiencies have been extracted for $\pi^0 p$ and $\pi^+ n$ reactions
and were found to be $\geq$90\%, in agreement with the simulation.
Since this paper deals with polarization observables, no special study was done
to assess the efficiency of multi track events.
Angular resolutions were also estimated via simulation, giving 3.5$^0$ in $\theta$
and 4.5$^0$ in $\varphi$.

Charged particle identification in the central region
is obtained by dE/dx technique thanks to a plastic scintillator barrel 
(32 bars, 5~mm thick, 43~cm long) with an energy resolution
$\approx$20\%. 
For the charged particles emitted in the forward
direction, a Time-of-Flight measurement is provided by a double plastic
scintillator hodoscope (300$\times$300$\times$3~cm$^3$) placed at a distance of 
3~m from the target and having a resolution of $\approx$600~ps. This detector provides 
also a measure of the energy loss dE/dx. Energy calibrations 
were extracted
from the analysis of the $\pi^0 p$ photoproduction reaction while
the ToF calibration of the forward wall was obtained from fast electrons
produced in the target.

Photons are detected in a BGO calorimeter made of 480 ($15 \theta \times 32 \varphi$)
crystals, each of 21 radiation lengths. They are identified
as clusters of adjacent crystals (3 on average for an energy threshold of 10~MeV per crystal)
with no associated hit in the barrel.
The measured energy resolution is 3\% on average ($E_\gamma$=200-1200~MeV). 
The angular resolution is 6$^0$ and 7$^0$ for polar and azimuthal
angles, respectively ($E_\gamma \geq$ 200~MeV and $l_{target}$=3~cm).

\section{Data analysis}
\label{analysis}

\subsection{Channel selection}
\label{event_sel}

For the present results, the charged decay of the $\Lambda$ ($\Lambda \rightarrow p\pi^-$, BR=63.9\%) 
was considered and the same selection method used in our previous publication on $K\Lambda$ photoproduction 
\cite{lle07} was applied. Only the main points will be recalled in the following.

Only events with three tracks and no neutral cluster detected in the BGO 
calorimeter were retained. In the absence of a direct 
measurement of energy and/or momentum of the charged particles,
the measured angles ($\theta$, $\varphi$) of the three tracks were combined with
kinematical constraints to calculate momenta. Particle identification was then
obtained from the association of the calculated momenta with dE/dx and/or ToF
measurements. 

The main source of background is the $\gamma p \rightarrow p \pi^+ \pi^-$
reaction, a channel with a similar final state and a cross section hundred times larger.
Selection of the $K\Lambda$ final state was achieved by applying narrow cuts on the following set of
experimental quantities:

\begin{itemize}

\item[.] Energy balance.
\vspace{0.3cm}

\item[.] Effective masses of the three particles 
extracted from the combination of measured dE/dx and ToF (only at forward angles) with
calculated momenta.
\vspace{0.3cm}

\item[.] Missing mass $m_{\gamma p- K^{+}}$ evaluated from
$E_\gamma$, $\theta_K$ (measured) and $p_K$ (calculated).
\vspace{0.3cm}

\end{itemize}

For each of these variables, the width $\sigma$ of the corresponding distribution
(Gaussian-like shape) were extracted from a Monte-Carlo simulation of the 
apparatus response based on the GEANT3 package of the CERN library. 

To check the quality of the event selection, the distribution of the $\Lambda$ decay length
was used due to its high sensitivity to background contamination.

\begin{figure}
\begin{center}
\includegraphics[width=0.9\linewidth]{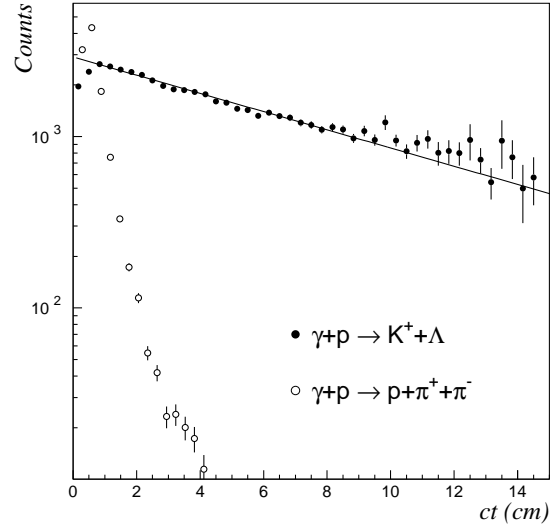} 
\end{center}
\caption{Reconstructed $\Lambda$ decay length spectrum after all selection cuts (closed circles)
for events with at least two tracks in the cylindrical chambers.
The solid line represents the fit with an 
exponential function $\alpha*exp(-ct/c\tau)$ where $\alpha$ and
$c\tau$ are free parameters. The second distribution 
(open circles) was obtained without applying selection cuts.
It corresponds to the main background reaction
($\gamma p \rightarrow p\pi^{+}\pi^{-}$) which, as expected, contributes
only to small ct values.}
\label{tfkl}
\end{figure}

Event by event, track information and $\Lambda$ momentum were combined to
obtain the distance $d$ between the reaction and decay vertices. 
The $\Lambda$ decay length
was then calculated with the usual formula $ct_{\Lambda}=d/(\beta_{\Lambda}*\gamma_{\Lambda})$.
Fig. \ref {tfkl} shows the resulting
distributions for events selected with all cuts at $\pm$2$\sigma$ (closed circles)
compared with events without cuts (open circles). 
These spectra were corrected
for detection efficiency losses estimated from the Monte-Carlo simulation
(significant only for ct$\ge$5~cm). It should be noted that the deficit in 
the first bins is attributed to finite resolution effects not fully taken 
into account in the simulation.

The first spectrum was fitted for ct$\geq$1~cm by an exponential function
$\alpha*exp(-ct/c\tau)$ with $\alpha$ and $c\tau$ as free parameters.
The fitted $c\tau$ value (8.17$\pm$0.31~cm) is in good agreement with the PDG expectation 
for the $\Lambda$ mean free path ($c\tau_\Lambda$=7.89~cm) \cite{pdg04}.

By contrast, the spectrum without cuts is dominated by $p \pi^+ \pi^-$ background
events. As expected, they contribute mostly to small ct values ($\le$2-3~cm),
making the shape of this distribution highly sensitive to background contamination.
For instance, a pronounced peak already shows up when opening selection cuts at 
$\pm$3$\sigma$.

A remaining source of background, which cannot be seen in the ct plot presented
above, originates from the contamination by the reaction 
$\gamma p \rightarrow K^+\Sigma^0$. Indeed, events where the decay photon is not detected
are retained by the first selection step. Since these events are kinematically
analyzed as $K\Lambda$ ones, most of them are nevertheless rejected by the selection cuts.
From the simulation, this contamination was found to be of the order of 2\%. 

\begin{figure}
\begin{center}
\includegraphics[width=0.9\linewidth]{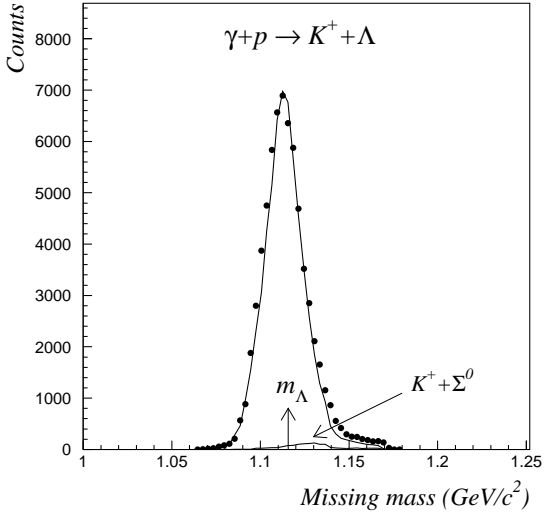} 
\end{center}
\caption{Distribution of the missing mass $m_{\gamma p- K^{+}}$ reconstructed from measured $E_\gamma$ and $\theta_K$ and calculated $p_K$.
Data after all selection cuts (closed circles) are compared to the simulation (solid line). 
The expected contribution from the reaction $\gamma p \rightarrow K^+\Sigma^0$
is also plotted (note that it is not centered on the $\Sigma^0$ mass due to kinematical constraints 
in the event analysis). 
The vertical arrow indicates the $\Lambda$ mass.}
\label{mkl}
\end{figure}

As a further check of the quality of the data sample, the missing mass spectrum
was calculated. One should remember that the missing mass is not directly 
measured and is not used as a criterion for the channel identification.
The spectrum presented in fig. \ref {mkl} (closed circles) is in fair
agreement with the simulated distribution (solid line). Some slight
discrepancies can nevertheless be seen in the high energy tail of the spectra.
The simulated missing mass distribution of the contamination from the
$\gamma p \rightarrow K^+\Sigma^0$ reaction,
also displayed in fig. \ref {mkl}, clearly indicates that
such a background cannot account for the observed differences. Rather,
these are attributed to the summation of a large number of data taking periods
with various experimental configurations (target
length, wire chambers, green vs UV laser line, ...). Although these
configurations were implemented in corresponding simulations,
small imperfections (misalignments in particular) could not be
taken into account.

To summarize, thanks to these experimental checks, we are confident that the level of 
background in our selected sample is limited. This is corroborated by the simulation from which the
estimated background contamination (multi-pions and $K^+\Sigma^0$ contributions) never exceeds 5\% 
whatever the incident photon energy or the meson recoil angle.

\subsection{Measurement of $O_x$, $O_z$ and $T$}

As will be shown below, the beam-recoil observables $O_x$ and $O_z$, as well as the target 
asymmetry $T$, can be extracted from the angular distribution of the $\Lambda$ decay proton.

\subsubsection{Formalism}
\label{expl}

For a linearly polarized beam and an unpolarized target, the differential cross section
can be expressed in terms of the single polarization observables $\Sigma$, $P$, $T$  
(beam asymmetry, recoil polarization, target asymmetry, respectively) and of the double polarization observables $O_x$,
$O_z$ (beam-recoil), as follows \cite{ade90}:

\noindent
\begin{eqnarray} 
\rho_f \frac{d\sigma}{d\Omega}&=& \frac{1}{2} \biggl (\frac{d\sigma}{d\Omega} \biggr)_{0}
[1 - P_{\gamma}\Sigma \cos 2\varphi_\gamma \nonumber \\
+\sigma_{x'} P_{\gamma} O_x \sin 2\varphi_{\gamma} \nonumber \\
+\sigma_{y'} (P - P_{\gamma} T \cos 2\varphi_{\gamma}) \nonumber \\
+\sigma_{z'} P_{\gamma} O_z \sin 2\varphi_{\gamma}]
\label{eq1}
\end{eqnarray}

\begin{figure}
\begin{center}
\includegraphics[width=0.9\linewidth]{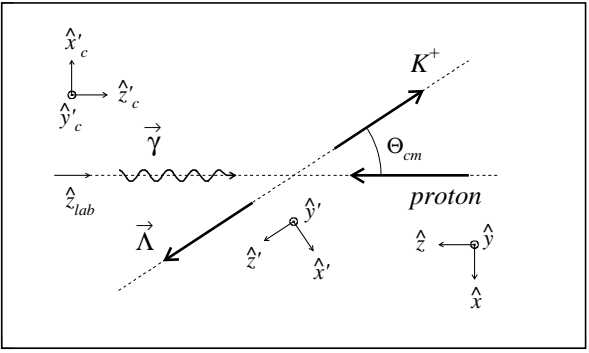} 
\end{center}
\caption{Definition of the coordinate systems and polar angles in the center-of-mass frame (viewed in the reaction plane).
The [$\hat{x}'$,$\hat{y}'$,$\hat{z}'$] system is used to specify the polarization of the outgoing $\Lambda$ baryon:
$\hat{z}'$ is along the $\Lambda$ momentum and $\hat{y}'$ perpendicular to the reaction plane.
The [$\hat{x}$,$\hat{y}$,$\hat{z}$] system is used to specify the incident photon polarization:
$\hat{z}$ is along the incoming proton momentum and $\hat{y}$ identical to $\hat{y}'$.
The polar angle $\theta_{cm}$ of the outgoing $K^+$ meson is defined with respect to the incident beam
direction $\hat{z}_{lab}$. [$\hat{x}'_c$,$\hat{y}'_c$,$\hat{z}'_c$] is the coordinate system chosen by the
CLAS collaboration for the $\Lambda$ polarization. The $\hat{x}'_c$ and $\hat{z}'_c$ axes are obtained from
$\hat{x}'$ and $\hat{z}'$ by a rotation of angle $\pi+\theta_{cm}$.}
\label{ax}
\end{figure}

\begin{figure}
\begin{center}
\includegraphics[width=0.9\linewidth]{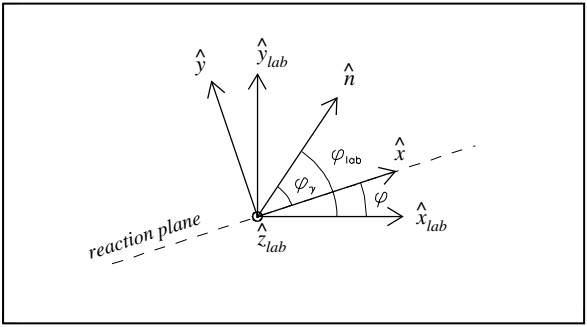} 
\end{center}
\caption{Definition of the coordinate systems and azimuthal angles in the center-of-mass frame (viewed perpendicularly to
the beam direction).
The [$\hat{x}_{lab}$,$\hat{y}_{lab}$,$\hat{z}_{lab}$] system corresponds to the laboratory axes with $\hat{z}_{lab}$
along the incident beam direction. 
The [$\hat{x}$,$\hat{y}$,$\hat{z}$] system, used to define the incident photon polarization,
has its axes $\hat{x}$ and $\hat{y}$ along and perpendicular to the reaction plane (azimuthal angle $\varphi$), respectively.
The polarization of the beam is along $\hat{n}$ (azimuthal angle $\varphi_{lab}$).
The two beam polarization states correspond to $\varphi_{lab}=0^0$ (horizontal) and $\varphi_{lab}=90^0$
(vertical) ($\varphi_{lab}=\varphi_{\gamma}+\varphi$).}
\label{ax2}
\end{figure}

\noindent
$\rho_f$ is the density
matrix for the lambda final state and $(d\sigma /d\Omega)_{0}$ the unpolarized differential cross section.
The Pauli matrices $\sigma_{x',y',z'}$ refer to the
lambda quantization axes defined by $\hat{z}'$ along the lambda momentum in the center-of-mass frame
and $\hat{y}'$ perpendicular to the reaction plane (fig. \ref{ax}).
$P_{\gamma}$ is the degree of linear polarization of the beam along an axis defined by
$\hat{n}=\hat{x}\cos \varphi_{\gamma}+\hat{y}\sin \varphi_{\gamma}$; the photon quantization axes are
defined by $\hat{z}$ along the proton center-of-mass momentum and $\hat{y}$=$\hat{y}'$ (fig. \ref{ax}).
We have $\varphi_{\gamma}
=\varphi_{lab}-\varphi$, where $\varphi_{lab}$ and $\varphi$ are the azimuthal angles of the
photon polarization vector and of the reaction plane in the laboratory axes, respectively (fig. \ref{ax2}).

The beam-recoil observables $C_x$ and $C_z$ measured by the CLAS collaboration with a 
circularly polarized beam \cite{bra07} were obtained
using another coordinate system for describing the hyperon polarization,
the $\hat{z}'$ axis being along the incident beam direction instead of the momentum of one of the recoiling
particles (see fig. \ref{ax}). Such a non-standard coordinate system was chosen to give the results their 
simplest interpretation in terms of polarization transfer but implied the model calculations to be adapted. 
To check the consistency of our results with the CLAS values (see sect. \ref{combi}), our $O_x$ and $O_z$ 
values were converted using the following rotation matrix:

\noindent
\begin{eqnarray}
O_x^c=-O_x \cos\theta_{cm}-O_z \sin\theta_{cm} \nonumber \\
O_z^c=O_x \sin\theta_{cm}-O_z \cos\theta_{cm}
\label{eqconv}
\end{eqnarray}

It should be noted that our definition for $O_x$ and $O_z$ (eq. \ref{eq1})
has opposite sign with respect to the definition given in the article \cite{bar75}, which is used in several hadronic models. 
We chose the same sign convention than the CLAS collaboration.

For an outgoing lambda with an arbitrary quantization axis $\hat{n}'$, the 
differential cross section becomes:

\noindent
\begin{eqnarray} 
\mathbf{P}_{\Lambda} \cdot \hat{n}' \frac{d\sigma}{d\Omega}&=&Tr \Big[ \mathbf{\sigma} \cdot \hat{n'} \rho_f \frac{d\sigma}{d\Omega} \Big]
\label{eq2}
\end{eqnarray}

\noindent
where $\mathbf{P}_{\Lambda}$ is the polarization vector of the lambda.
If the polarization is not observed, the expression for the differential cross section reduces to:

\noindent
\begin{eqnarray} 
\frac{d\sigma}{d\Omega}&=&Tr \Big[\rho_f \frac{d\sigma}{d\Omega} \Big]
\label{eq3}
\end{eqnarray}

\noindent
which leads to:

\noindent
\begin{eqnarray} 
\frac{d\sigma}{d\Omega}&=& \biggl (\frac{d\sigma}{d\Omega} \biggr)_{0}
 [1 - P_{\gamma}\Sigma \cos 2\varphi_\gamma]
\label{eq4}
\end{eqnarray}

\noindent
For horizontal ($\varphi_{lab}=0^0$) and vertical ($\varphi_{lab}=90^0$) photon polarizations,
the corresponding azimuthal distributions of the reaction plane are therefore:

\noindent
\begin{eqnarray} 
\frac{d\sigma}{d\Omega}(\varphi_{lab}=0^0)&=& \biggl (\frac{d\sigma}{d\Omega} \biggr)_{0}
 [1 - P_{\gamma}\Sigma \cos 2\varphi]
\label{eq5}
\end{eqnarray}

\noindent
\begin{eqnarray} 
\frac{d\sigma}{d\Omega}(\varphi_{lab}=90^0)&=& \biggl (\frac{d\sigma}{d\Omega} \biggr)_{0}
 [1 + P_{\gamma}\Sigma \cos 2\varphi]
\label{eq6}
\end{eqnarray}

\noindent
The beam asymmetry values $\Sigma$ published in \cite{lle07} were extracted from the fit of the azimuthal
distributions of the ratio:
\noindent
\begin{eqnarray} 
\frac{N(\varphi_{lab}=90^0)-N(\varphi_{lab}=0^0)}{N(\varphi_{lab}=90^0)+N(\varphi_{lab}=0^0)}=P_\gamma \Sigma \cos 2\varphi
\label{eqsig}
\end{eqnarray}
\noindent

\subsubsection{$\Lambda$ polarization and spin observables}
\label{polo}

The components of the lambda polarization vector deduced from
eqs. \ref{eq1} to \ref{eq4} are:

\noindent
\begin{eqnarray} 
P_{\Lambda}^{x',z'}&=&\frac{P_{\gamma} O_{x,z} \sin 2\varphi_{\gamma}}{1 - P_{\gamma} \Sigma \cos 2\varphi_{\gamma}} 
\end{eqnarray}

\noindent
\begin{eqnarray} 
P_{\Lambda}^{y'}&=&\frac{P - P_{\gamma} T \cos 2\varphi_{\gamma}}{1 - P_{\gamma} \Sigma \cos 2\varphi_{\gamma}}
\end{eqnarray}

\noindent
These equations provide the connection between the $\Lambda$ polarisation $\mathbf{P}_{\Lambda}$
and the spin observables $\Sigma$, $P$, $T$, $O_x$ and $O_z$.

Integration of the polarization components over the azimuthal angle $\varphi$ of the reaction plane writes:

\noindent
\begin{eqnarray} 
<P_{\Lambda}^{i}>&=&\frac{\int P_{\Lambda}^{i}(\varphi)\frac{d\sigma}{d\Omega}(\varphi)d\varphi}
{\int \frac{d\sigma}{d\Omega}(\varphi)d\varphi}
\end{eqnarray}

\noindent
where $i$ stands for $x'$, $y'$ or $z'$.

When integrating over the full angular domain, the averaged $x'$ and $z'$
components of the polarization vector vanish while the $y'$ component is equal to $P$.
On the other hand, when integrating over appropriatly chosen angular sectors, all three averaged 
components can remain different from zero. For horizontal and vertical beam polarizations,
the following expressions are obtained when considering the four particular $\varphi$ domains 
defined hereafter \cite{cal97} (recalling $\varphi_{\gamma}=\varphi_{lab}-\varphi$):

\begin{itemize}

\item[.] $S_1 = [\pi /4,3\pi /4] \cup [5\pi /4,7\pi /4]$: \\
$<P_{\Lambda}^{y'}>(\varphi_{lab}=0^0)=(P\pi + 2P_{\gamma} T) / (\pi + 2P_{\gamma} \Sigma)$ \\
$<P_{\Lambda}^{y'}>(\varphi_{lab}=90^0)=(P\pi - 2P_{\gamma} T) / (\pi - 2P_{\gamma} \Sigma)$
\vspace{0.3cm}

\item[.] $S_2 = [-\pi /4,\pi /4] \cup [3\pi /4,5\pi /4]$: \\
$<P_{\Lambda}^{y'}>(\varphi_{lab}=0^0)=(P\pi - 2P_{\gamma} T) / (\pi - 2P_{\gamma} \Sigma)$ \\
$<P_{\Lambda}^{y'}>(\varphi_{lab}=90^0)=(P\pi + 2P_{\gamma} T) / (\pi + 2P_{\gamma} \Sigma)$
\vspace{0.3cm}

\item[.] $S_3 = [0,\pi /2] \cup [\pi ,3\pi /2]$: \\
$<P_{\Lambda}^{x',z'}>(\varphi_{lab}=0^0)=-2 P_{\gamma} O_{x,z} / \pi$ \\
$<P_{\Lambda}^{x',z'}>(\varphi_{lab}=90^0)=+2 P_{\gamma} O_{x,z} / \pi$
\vspace{0.3cm}

\item[.] $S_4 = [\pi /2,\pi] \cup [3\pi /2 ,2\pi]$ : \\
$<P_{\Lambda}^{x',z'}>(\varphi_{lab}=0^0)=+2 P_{\gamma} O_{x,z} / \pi$ \\
$<P_{\Lambda}^{x',z'}>(\varphi_{lab}=90^0)=-2 P_{\gamma} O_{x,z} / \pi$
\vspace{0.3cm}

\end{itemize}

\noindent
It should be noted that these four sectors cover the full $\varphi$ range.
In the following, these different combinations of $\varphi$ sectors and polarization states
will be labelled by the sign plus or minus appearing in the corresponding expressions for $<P_{\Lambda}^{i}>$.

\subsubsection{Decay angular distribution}

In the lambda rest frame, the angular distribution of the decay proton is given by \cite{lee57}:

\noindent
\begin{eqnarray} 
W(\cos\theta_{p})=\frac{1}{2} \big(1+\alpha |\mathbf{P}_{\Lambda}| \cos\theta_{p} \big)
\label{dist_ang}
\end{eqnarray} 

\noindent
where $\alpha$=0.642$\pm$0.013 \cite{pdg04} is the $\Lambda$ decay parameter and $\theta_{p}$ the
angle between the proton direction and the lambda polarization vector.

From this expression, one can derived an angular distribution for each 
component of $\mathbf{P}_{\Lambda}$:

\noindent
\begin{eqnarray} 
W(\cos\theta_{p}^i)=\frac{1}{2} \big(1+\alpha P_{\Lambda}^{i} \cos\theta_{p}^i \big)
\label{dist_ang2}
\end{eqnarray}

\noindent
where $\theta_{p}^i$ is now the angle between the proton direction and the quantization axis $i$ ($x'$, $y'$ or $z'$).

The components being
determined in the $\Lambda$ rest frame, a suitable transformation should be applied to calculate them
in the center-of-mass frame. However, as the boost direction is along the lambda momentum,
it can be shown that the polarization measured in the lambda rest frame remains unchanged in the
center-of-mass frame \cite{bra07}.

When integrating over all possible azimuthal angles $\varphi$, the proton angular distribution
with respect to the $y'$-axis simply writes:

\noindent
\begin{eqnarray} 
W(\cos\theta_{p}^{y'})=\frac{1}{2} (1 + \alpha P \cos\theta_{p}^{y'})
\label{eqwyp}
\end{eqnarray}

\noindent
where $P$ is the recoil polarization.
Our $P$ results published in \cite{lle07} were determined directly from the measured up/down
asymmetry:

\noindent
\begin{eqnarray}
\frac{N(\cos\theta_{p}^{y'}>0)-N(\cos\theta_{p}^{y'}<0)}{N(\cos\theta_{p}^{y'}>0)+N(\cos\theta_{p}^{y'}<0)}=\frac{1}{2} \alpha P
\label{eqp}
\end{eqnarray}

When integrating over the different angular domains specified above (sectors $S_1+S_2$ for $y'$-axis and 
$S_3+S_4$ for $x'$-,$z'$-axes, appropriatly combined with the two beam polarization),
the proton angular distributions with respect to 
the three quantization axes can be written as follows:

\noindent
\begin{eqnarray} 
W_\pm(\cos\theta_{p}^{x',z'})=\frac{1}{2} \big(1 \pm \alpha \frac{2P_{\gamma}O_{x,z}}{\pi} \cos\theta_{p}^{x',z'} \big)
\label{eqwxz}
\end{eqnarray}

\noindent
\begin{eqnarray} 
W_\pm(\cos\theta_{p}^{y'})=\frac{1}{2} \big(1 + \alpha \frac{P\pi \pm 2P_{\gamma}T}{\pi \pm 2P_{\gamma}\Sigma} \cos\theta_{p}^{y'} \big)
\label{eqwy}
\end{eqnarray}

\subsubsection{Experimental extraction}

As for $\Sigma$ and $P$, the observables $O_x$, $O_z$ and $T$ were extracted from ratios of the angular
distributions, in order to get rid of most of the distorsions introduced 
by the experimental acceptance.

Including the detection efficiencies, the yields measured as a function of
the proton angle with respect to the different axes write:

\noindent
\begin{eqnarray} 
N_\pm^{x',z'}=\frac{1}{2} N_{0\pm}^{x',z'} \epsilon_\pm (\cos\theta_{p}^{x',z'}) \big(1 \pm \alpha \frac{2P_{\gamma}O_{x,z}}{\pi} \cos\theta_{p}^{x',z'} \big)\nonumber \\
\label{eqnx}
\end{eqnarray}

\noindent
\begin{eqnarray} 
N_\pm^{y'}=\frac{1}{2}N_{0\pm}^{y'} \epsilon_\pm (\cos\theta_{p}^{y'}) \big(1 + \alpha \frac{P\pi \pm 2P_{\gamma}T}{\pi \pm 2P_{\gamma}\Sigma} \cos\theta_{p}^{y'} \big)
\label{eqny1}
\end{eqnarray}

\noindent
From the integration of the azimuthal distributions given by eqs. \ref{eq5} and \ref{eq6} over the different angular sectors,
it can be shown that:

\noindent
\begin{eqnarray} 
N_{0+}^{x',z'}=N_{0-}^{x',z'}
\label{eq7}
\end{eqnarray} 

\noindent
\begin{eqnarray} 
\frac{N_{0+}^{y'}}{N_{0-}^{y'}}=\frac{\pi + 2P_{\gamma}\Sigma}{\pi - 2P_{\gamma}\Sigma}
\label{eq8}
\end{eqnarray} 

\noindent
Assuming that the detection efficiencies do not depend on the considered $\varphi$ sectors
($\epsilon_+(\cos\theta_{p}^{i})=\epsilon_-(\cos\theta_{p}^{i})$ -  
the validity of this assumption will be discussed later on), we can then calculate
the following sums and ratios from which the efficiency cancels out:

\noindent
\begin{eqnarray}
N_+^{x',z'} + N_-^{x',z'} = \frac{1}{2} (N_{0+}^{x',z'}+N_{0-}^{x',z'})\epsilon_\pm (\cos\theta_{p}^{x',z'})\nonumber \\
\label{eqnxz}
\end{eqnarray}

\noindent
\begin{eqnarray}
N_+^{y'} + N_-^{y'} = \frac{1}{2} (N_{0+}^{y'}+N_{0-}^{y'})\epsilon_\pm (\cos\theta_{p}^{y'})(1 + \alpha P \cos\theta_{p}^{y'})\nonumber \\
\label{eqny}
\end{eqnarray}

\noindent
\begin{eqnarray} 
\frac {2 N_+^{x',z'}}{N_+^{x',z'}+ N_-^{x',z'}}=(1+\alpha \frac{2P_{\gamma}O_{x,z}}{\pi} \cos\theta_{p}^{x',z'})
\label{eqrxz}
\end{eqnarray}

\noindent
\begin{eqnarray} 
\frac {2 N_+^{y'}}{N_+^{y'}+ N_-^{y'}}=\big( 1+\frac{2P_\gamma \Sigma}{\pi}\big) \big( \frac{1 + \alpha \frac{P\pi+2P_{\gamma}T}{\pi+2P_{\gamma}\Sigma} \cos\theta_{p}^{y'}}{1+\alpha P \cos\theta_{p}^{y'}} \big)
\label{eqry}
\end{eqnarray}

\begin{figure}
\begin{center}
\includegraphics[width=1.0\linewidth]{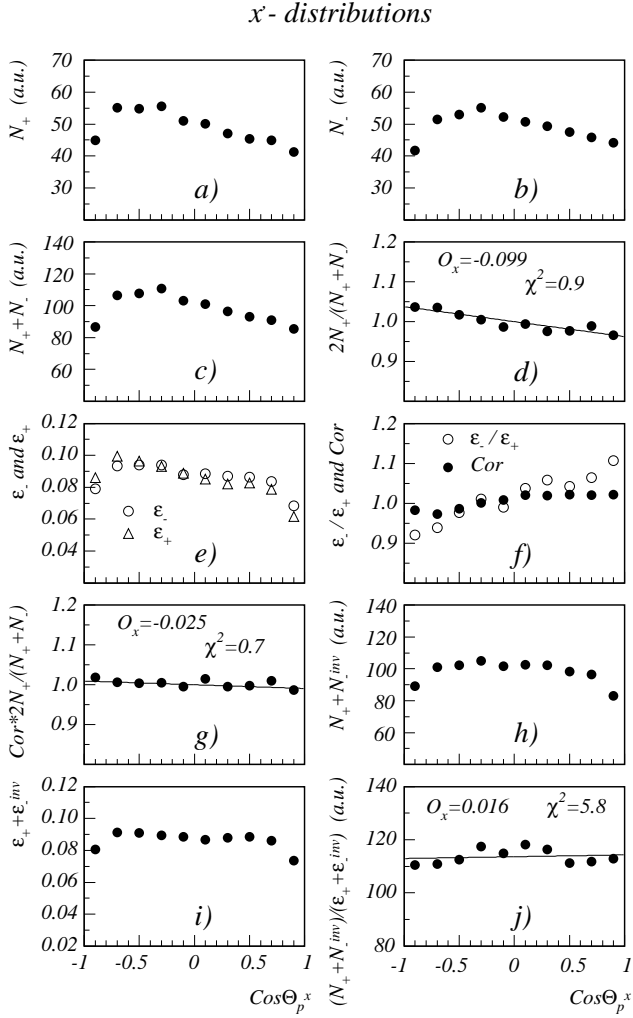} 
\end{center}
\caption{Angular distributions for the decay proton in the lambda rest frame with respect to the $x'$-axis:
a) distribution $N_+$;
b) distribution $N_-$; 
c) sum $N_++N_-$; 
d) ratio $2N_+/(N_++N_-)$; 
e) efficiencies $\epsilon_+$ (triangles) and $\epsilon_-$ (circles) calculated from the simulation;
f) ratio $\epsilon_-/\epsilon_+$ (open circles) and correction factor $Cor$ (closed circles) given by eq. \ref{rcor} calculated from the simulation;
g) ratio $2N_+/(N_++N_-)$ corrected by the factor $Cor$;
h) distribution $N_++N_-^{inv}$, with $N^{inv}=N(-\cos\theta_{p})$; 
i) efficiency $\epsilon_++\epsilon_-^{inv}$, with $\epsilon^{inv}=\epsilon(-\cos\theta_{p})$, calculated from the simulation;
j) distribution $N_++N_-^{inv}$ corrected by the efficiency $\epsilon_++\epsilon_-^{inv}$ .
The solid line in d) and g) represents the fit by the (linear) function given in the r.h.s. of eq. \ref{eqrxz}.
The solid line in j) represents the fit by the (linear) function given in the r.h.s. of eq. \ref{eqfxz}.
The reduced-$\chi^2$ and the $O_x$ value obtained from the fits are reported in d), g) and j).}
\label{ox_fit}
\end{figure}

\begin{figure}
\begin{center}
\includegraphics[width=1.0\linewidth]{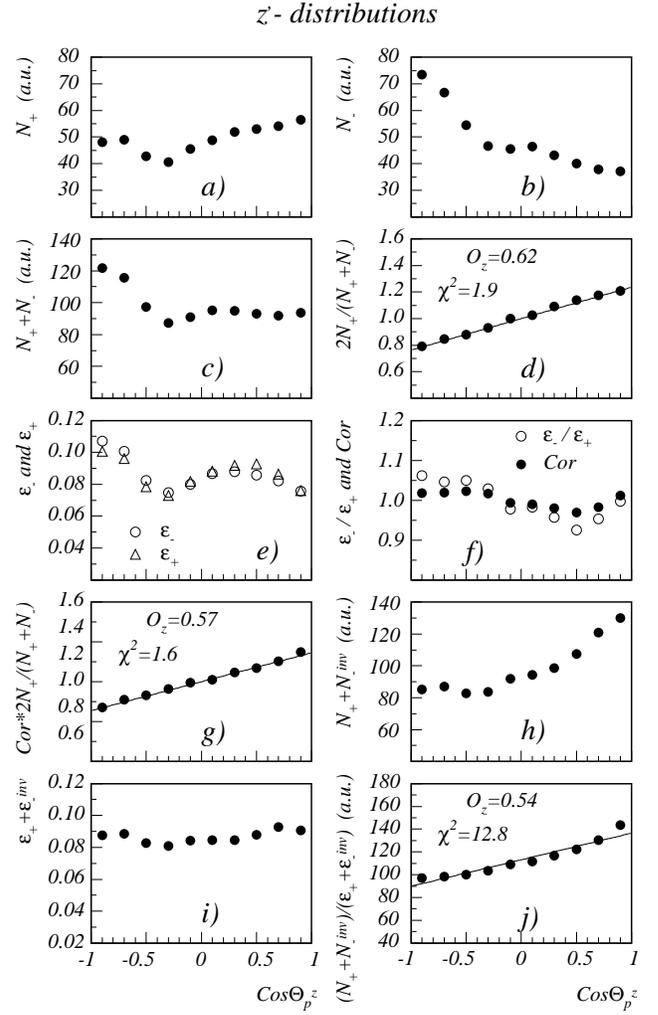} 
\end{center}
\caption{Angular distributions for the decay proton in the lambda rest frame with respect to the $z'$-axis
(all distributions as in fig. \ref{ox_fit}).
The reduced-$\chi^2$ and the $O_z$ value obtained from the fits are reported in d), g) and j).}
\label{oz_fit}
\end{figure}

\begin{figure}
\begin{center}
\includegraphics[width=1.0\linewidth]{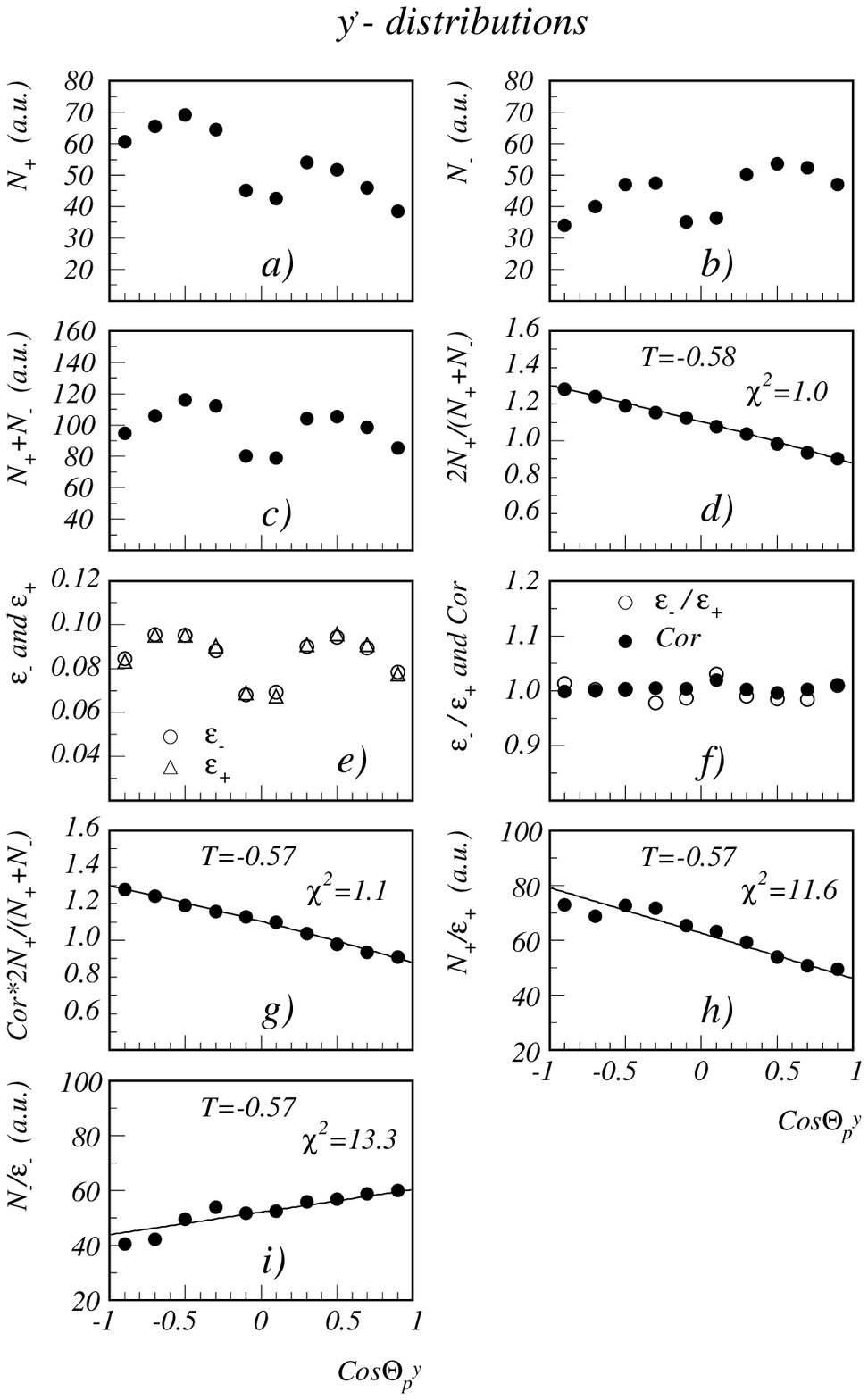} 
\end{center}
\caption{Angular distributions for the decay proton in the lambda rest frame with respect to the $y'$-axis:
a) distribution $N_+$;
b) distribution $N_-$; 
c) sum $N_++N_-$; 
d) ratio $2N_+/(N_++N_-)$; 
e) efficiencies $\epsilon_+$ (triangles) and $\epsilon_-$ (circles) calculated from the simulation - they are
symmetrical about $\theta_{cm}=90^0$ (we find $\epsilon_{down}/\epsilon_{up}$=1.03);
f) ratio $\epsilon_-/\epsilon_+$ (open circles) and correction factor $Cor$ (closed circles) given by eq. \ref{rcor} calculated from the simulation;
g) ratio $2N_+/(N_++N_-)$ corrected by the factor $Cor$;
h) distribution $N_+$ corrected by the efficiency $\epsilon_+$;
i) distribution $N_-$ corrected by the efficiency $\epsilon_-$.
The solid line in d) and g) represents the fit by the (non-linear) function given in the r.h.s. of eq. \ref{eqry}.
These distributions exhibit a linear behaviour since the overall recoil polarisation P is very low (the value
extracted from the up/down asymmetry of the raw distribtion $N_++N_-$ is -0.12).
The solid line in h) and i) represents the simultaneous fit by the (linear) functions given in the r.h.s. of eqs. \ref{eqfyp} and \ref{eqfym}.
The reduced-$\chi^2$ and the $T$ value obtained from the fits are reported in d), g), h) and i).}
\label{t_fit}
\end{figure}

To illustrate the extraction method of $O_{x}$, $O_{z}$ and $T$, the $N_+$ and $N_-$ experimental
distributions together with their sums and ratios,
summed over all photon energies and meson polar angles, 
are displayed in figs. \ref{ox_fit} ($x'$-axis), \ref{oz_fit} ($z'$-axis) and \ref{t_fit} ($y'$-axis). 
Thanks to the efficiency correction given by the distributions
$N_++N_-$ (figs. \ref{ox_fit},\ref{oz_fit},\ref{t_fit}-c), the ratios $2N_+/(N_++N_-)$ (figs. \ref{ox_fit},\ref{oz_fit},\ref{t_fit}-d), 
from which the efficiency drops out, exhibit the expected dependence
in $\cos\theta_{p}$ and can be therefore fitted by the functions given in the r.h.s. of eqs. \ref{eqrxz} and \ref{eqry}. 
The known energy dependence of $P_{\gamma}$ and the previously measured values for
$\Sigma$ and $P$ \cite{lle07} are then used to deduce $O_{x}$, $O_{z}$ and $T$ from
the fitted slopes.

The validity of the hypothesis $\epsilon_+(\cos\theta_{p}^{i})=\epsilon_-(\cos\theta_{p}^{i})$ was studied via the Monte Carlo 
simulation in which a polarized $\Lambda$ decay was included. 
The efficiencies $\epsilon_\pm$ calculated from the simulation are presented in plots e) of
figs. \ref{ox_fit} to \ref{t_fit} and the ratios $\epsilon_-/\epsilon_+$ in plots f) (open circles). 
As one can see, for the $y'$ case, this ratio remains very close to 1 whatever the angle while, for $x'$ and $z'$, 
the discrepancy from 1 is more pronounced and evolves with the angle. This shows that some corrections
should be applied on the measured ratios $2N_+/(N_++N_-)$ to take into account the non-negligible differences
observed between $\epsilon_+$ and $\epsilon_-$. The correction factors, plotted in figs. \ref{ox_fit},\ref{oz_fit},\ref{t_fit}-f) 
(closed circles), were calculated through the following expression:

\noindent
\begin{eqnarray}
Cor=\big( \frac {2 N_+^i}{N_+^i+ N_-^i} \big)_{gen}/\big( \frac {2 N_+^i}{N_+^i+ N_-^i} \big)_{sel}
\label{rcor}
\end{eqnarray}

\noindent
where {\it gen} and {\it sel} stand for generated and selected events.
Since $\epsilon_\pm=(N_\pm)_{sel}/(N_\pm)_{gen}$, it can be re-written as:

\noindent
\begin{eqnarray}
Cor=\frac{1}{2}\big( \frac {2 N_+^i}{N_+^i+ N_-^i} \big)_{gen} [1+\frac{\epsilon_-^i}{\epsilon_+^i} \big( \frac {N_-^i}{N_+^i} \big)_{gen}]
\label{rcor2}
\end{eqnarray}

\noindent
The corrected distributions are displayed in the plots g) of figs. \ref{ox_fit} to \ref{t_fit}. 
After correction, as expected, the slope of the $y'$ distribution is unaffected while the slopes of the $x'$ and $z'$ 
distributions are slightly modified. These distributions were again fitted 
to obtain the final values of $O_{x}$, $O_{z}$ and $T$.

As the detection efficiencies and the correction factors calculated from
the simulation depend on the input values of $O_{x}$, $O_{z}$ and $T$, an iterative method
was used. Three iterations were sufficient to reach stable values.

For a consistency check, an alternative extraction method was implemented.
The angular distributions were directly corrected by the simulated efficiencies 
and fitted according to:

\noindent
\begin{eqnarray}
\frac{N_+^{x',z'}+N_-^{x',z',inv}} {\epsilon_+^{x',z'}+\epsilon_-^{x',z',inv}}=\frac{1}{2} N_{0+}^{x',z'}\big(1+\alpha \frac{2P_{\gamma}O_{x,z}}{\pi} \cos\theta_{p}^{x',z'} \big)\nonumber \\
\label{eqfxz}
\end{eqnarray}

\noindent
\begin{eqnarray}
\frac{N_+^{y'}} {\epsilon_+^{y'}}=\frac{1}{2} N_{0+}^{y'}\big(1+\alpha \frac{P\pi+2P_{\gamma}T}{\pi+2P_{\gamma}\Sigma} \cos\theta_{p}^{y'} \big)
\label{eqfyp}
\end{eqnarray}

\noindent
\begin{eqnarray}
\frac{N_-^{y'}} {\epsilon_-^{y'}}=\frac{1}{2} N_{0+}^{y'}\frac{\pi-2P_{\gamma}\Sigma}{\pi+2P_{\gamma}\Sigma}\big(1+\alpha \frac{P\pi-2P_{\gamma}T}{\pi-2P_{\gamma}\Sigma} \cos\theta_{p}^{y'} \big)
\label{eqfym}
\end{eqnarray}

\noindent
where $N^{inv}$ and $\epsilon^{inv}$ stand for $N(-\cos\theta_{p})$ and $\epsilon(-\cos\theta_{p})$, respectively.
This trick, used for the $x'$ and $z'$ cases, allows to combine the $N_+$ and $N_-$ distributions which have opposite slopes
(eq. \ref{eqnx}).

To illustrate this second extraction method, the corrected distributions, summed over all photon energies
and meson polar angles, are displayed in figs. \ref{ox_fit},\ref{oz_fit}-j) ($x',z'$-axes) and \ref{t_fit}-h),i) ($y'$-axis).
They were obtained by dividing the originally measured distributions (figs. \ref{ox_fit},\ref{oz_fit}-h and
\ref{t_fit}-a,b) by the corresponding efficiency distributions (figs. \ref{ox_fit},\ref{oz_fit}-i and \ref{t_fit}-e).
In the $y'$-axis case, the two corrected spectra $N_\pm/\epsilon_\pm$ were simultaneously fitted.

This method gives results in good agreement with those extracted from the first method.
Nevertheless, the resulting $\chi^2$ were found to be significantly larger 
(the global reduced-$\chi^2$ values are given in figs. \ref{ox_fit} to \ref{t_fit} - they are close to 1 for the first method
and five to ten times larger for the second one). The first method, which
relies upon ratios leading to an intrinsic first order efficiency correction, is
less dependent on the simulation details and was therefore preferred.

Three sources of systematic errors were taken into account: the laser beam polarization ($\delta P_\gamma / P_\gamma$=2\%),
the $\Lambda$ decay parameter $\alpha$ ($\delta\alpha=0.013$) and the hadronic background.
The error due to the hadronic background was estimated from the variation
of the extracted values when cuts were changed from $\pm$2$\sigma$ to $\pm$2.5$\sigma$.
Given the good agreement between the two extraction methods, no corresponding systematic error was considered.
For the $T$ observable, the measured values for
$\Sigma$ and $P$ being involved, their respective errors were included in the estimation of the uncertainty.
All systematic and statistical errors have been summed quadratically.

\section{Results and discussions}

The complete set of beam-recoil polarization and target asymmetry data is displayed in figs. \ref{oxkl} to \ref{pcokl}. These data
cover the production threshold region ($E_\gamma$=911-1500 MeV) and a large angular range ($\theta_{cm}^{kaon}=30-140^0$).
Numerical values are listed in tables \ref{table_oxkl} to \ref{table_tkl}. Error bars are the quadratic sum of statistical
and systematic errors.

\subsection{Observable combination and consistency check}
\label{combi}

In pseudoscalar meson photoproduction, one can extract experimentally 16 different quantities: the 
unpolarized differential cross section $(d\sigma/d\Omega)_0$,
3 single polarization observables ($P$, $T$, $\Sigma$), 4 beam-target polarizations ($E$, $F$, $G$, $H$), 4 beam-recoil polarizations
($C_x$, $C_z$, $O_x$, $O_z$) and 4 target-recoil polarizations ($T_x$, $T_z$, $L_x$, $L_z$).
The various spin observables are not independent but are constrained by non-linear identities and various
inequalites \cite{ade90}, \cite{bar75}, \cite{chi97}, \cite{art07}. 
In particular, of the seven single and beam-recoil polarization observables, only five are independent being
related by the two equations:

\noindent
\begin{eqnarray}
C_x^2+C_z^2+O_x^2+O_z^2=1+T^2-P^2-\Sigma^2
\label{eqobs1}
\end{eqnarray}

\noindent
\begin{eqnarray}
C_z O_x-C_x O_z=T- P \Sigma
\label{eqobs1b}
\end{eqnarray}

\noindent
There are also a number of inequalities involving three of these observables:

\noindent
\begin{eqnarray}
|T \pm P| \leq 1 \pm \Sigma
\label{eqobs6}
\end{eqnarray}

\noindent
\begin{eqnarray}
P^2+O_x^2+O_z^2 \leq 1
\label{eqobs2}
\end{eqnarray}

\noindent
\begin{eqnarray}
\Sigma^2+O_x^2+O_z^2 \leq 1
\label{eqobs3}
\end{eqnarray}

\noindent
\begin{eqnarray}
P^2+C_x^2+C_z^2 \leq 1
\label{eqobs4}
\end{eqnarray}

\noindent
\begin{eqnarray}
\Sigma^2+C_x^2+C_z^2 \leq 1
\label{eqobs5}
\end{eqnarray}

These different identities and inequalities can be used to test the consistency of our present and previous measurements.
They can also be used to check the compatibility of our data with the results on $C_x$ and $C_z$ recently published by 
the CLAS collaboration \cite{bra07}. 

\begin{figure}
\begin{center}
\includegraphics[width=0.9\linewidth]{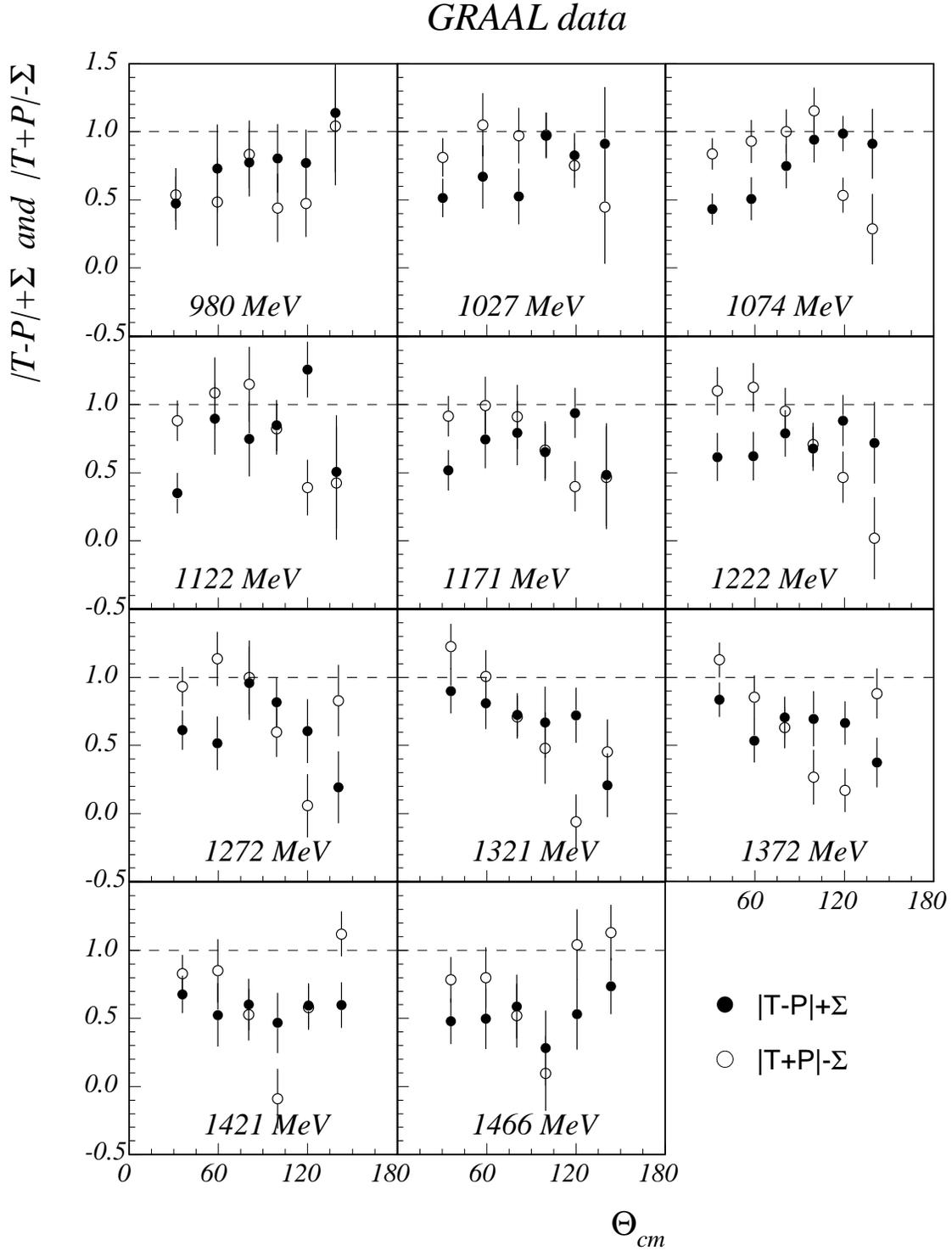} 
\end{center}
\caption{Angular distributions of the quantities $|T-P|+\Sigma$ (closed circles) and $|T+P|-\Sigma$ (open circles). 
We should have the inequalities $|T\pm P|\mp \Sigma \leq 1$ (eq. \ref{eqobs6}).}
\label{tpskl}
\end{figure}

\begin{figure}
\begin{center}
\includegraphics[width=0.9\linewidth]{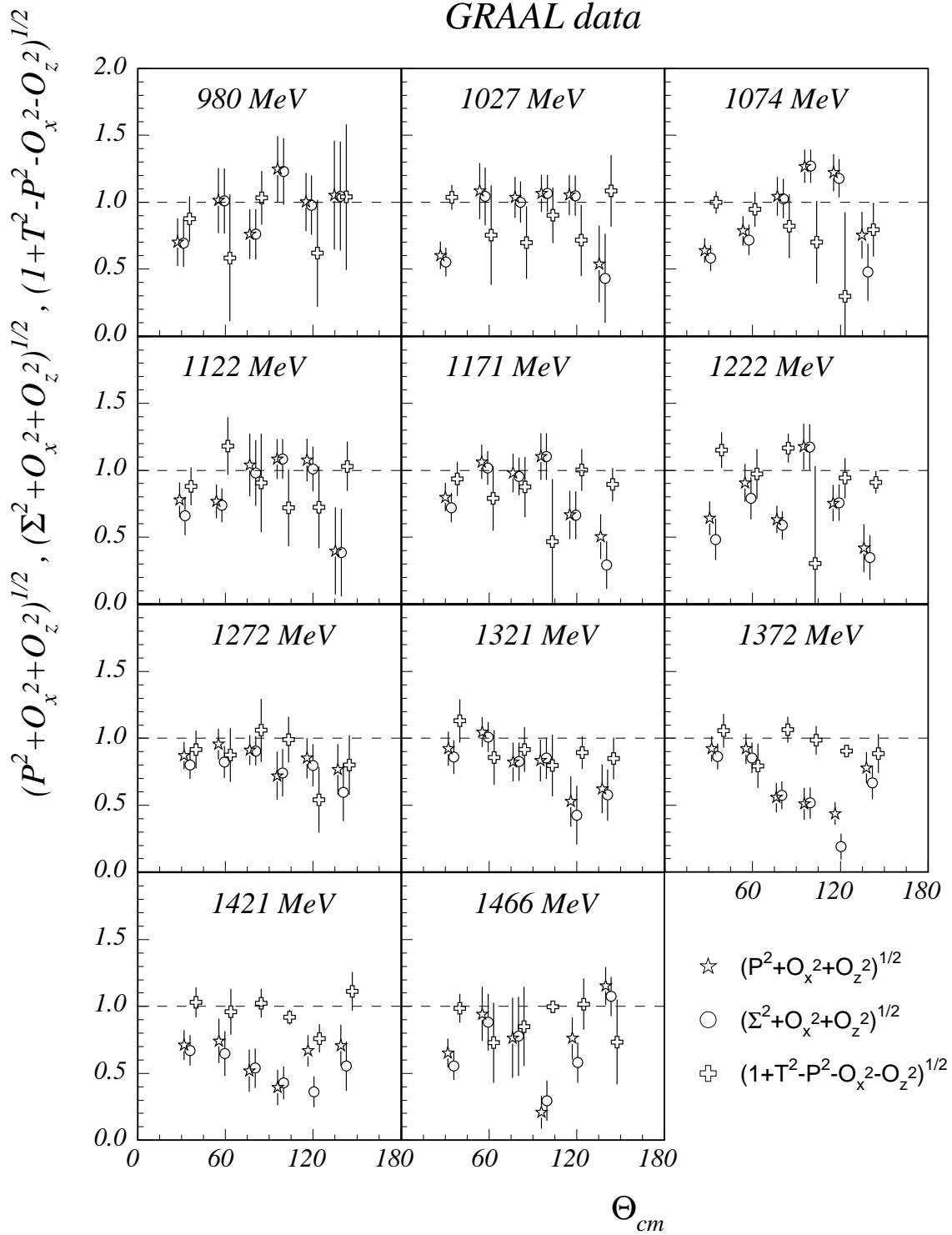} 
\end{center}
\caption{Angular distributions of the quantities $(P^2+O_x^2+O_z^2)^{1/2}$ (stars), $(\Sigma^2+O_x^2+O_z^2)^{1/2}$ (circles) and
$(1+T^2-P^2-O_x^2-O_z^2)^{1/2}=(\Sigma^2+C_x^2+C_z^2)^{1/2}$ (crosses). The first and third sets of data are horizontally shifted for visualization.
All these quantities should be $\leq 1$ (inequalities \ref{eqobs2}, \ref{eqobs3} and \ref{eqobs5}).}
\label{psockl}
\end{figure}

\begin{figure}
\begin{center}
\includegraphics[width=0.9\linewidth]{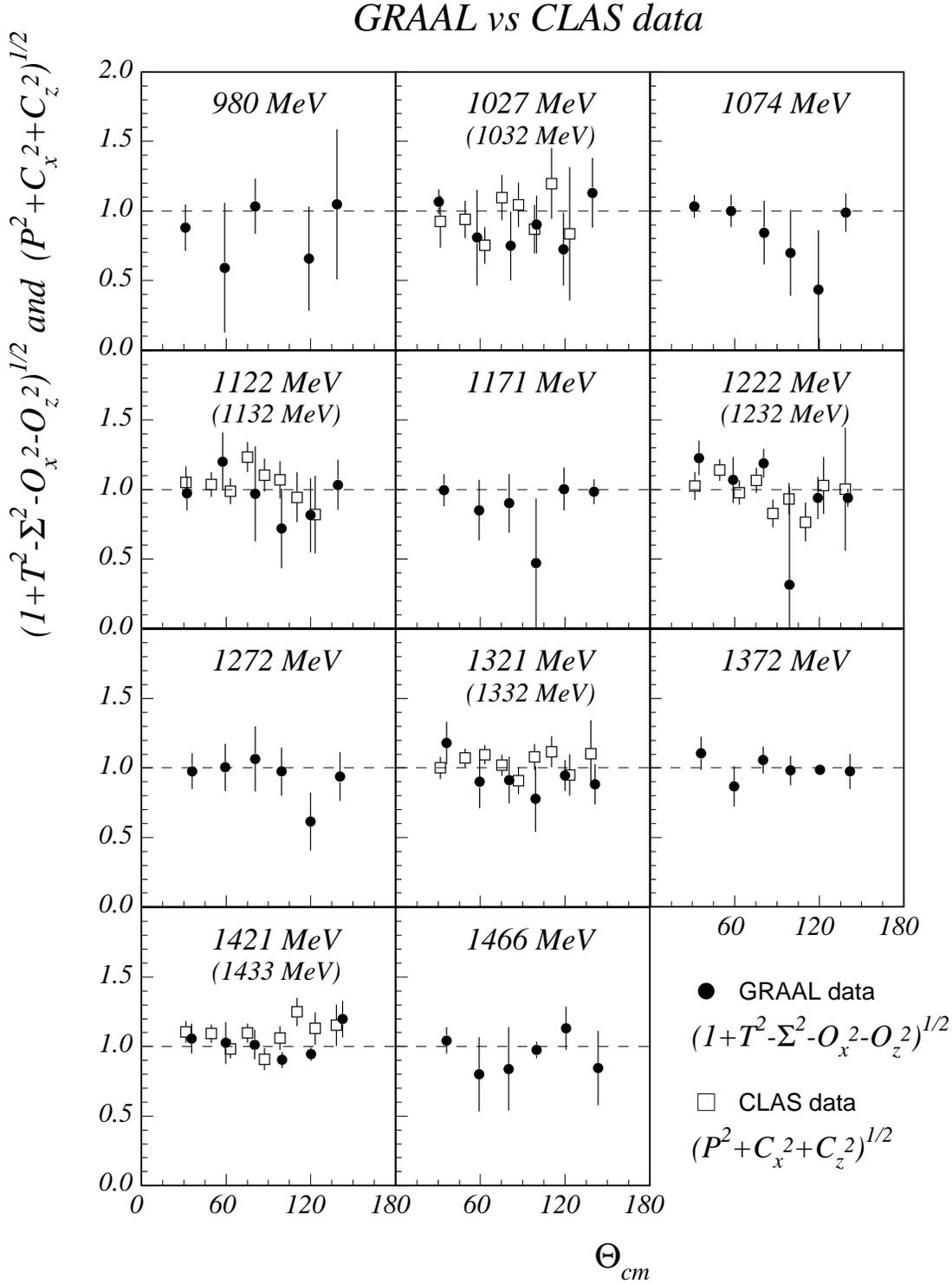} 
\end{center}
\caption{Angular distributions of the quantity $(1+T^2-\Sigma^2-O_x^2-O_z^2)^{1/2}=(P^2+C_x^2+C_z^2)^{1/2}$. 
This quantity should be $\leq 1$ (inequality \ref{eqobs4}).
Comparison to the values $(P^2+C_x^2+C_z^2)^{1/2}$ published by the CLAS collaboration (open squares - energy in parentheses).
Note that the $O_x^2+O_z^2$ and $C_x^2+C_z^2$ values are independent of the choice for the axes
specifying the $\Lambda$ polarization (see sect. \ref{expl}).}
\label{pckl}
\end{figure}

Our measured values for $\Sigma$, $P$, $T$, $O_x$ and $O_z$ were combined to test the above inequalities. 
Equation \ref{eqobs1} was used to calculate the quantity $C_x^2+C_z^2$ appearing in expressions \ref{eqobs4} and \ref{eqobs5}.
The results for the two combinations $|T\pm P|\mp \Sigma$ of the three single polarizations are
presented in fig. \ref{tpskl}.
The results for the quantities:

\begin{itemize}

\item[.] $(P^2+O_x^2+O_z^2)^{1/2}$, 
\item[.] $(\Sigma^2+O_x^2+O_z^2)^{1/2}$,
\item[.] $(1+T^2-P^2-O_x^2-O_z^2)^{1/2} = (\Sigma^2+C_x^2+C_z^2)^{1/2}$,
\item[.] $(1+T^2-\Sigma^2-O_x^2-O_z^2)^{1/2} = (P^2+C_x^2+C_z^2)^{1/2}$, 

\end{itemize}

\noindent
which combine single and double polarization observables, are displayed in figs. \ref{psockl} and \ref{pckl}. 
All these quantities should be $\leq 1$. The plotted uncertainties are given by the standard error
propagation. Whatever the photon energy or the meson polar angle, 
no violation of the expected inequalities is observed, confirming the internal consistency of our set of data.

\begin{figure}
\begin{center}
\includegraphics[width=0.9\linewidth]{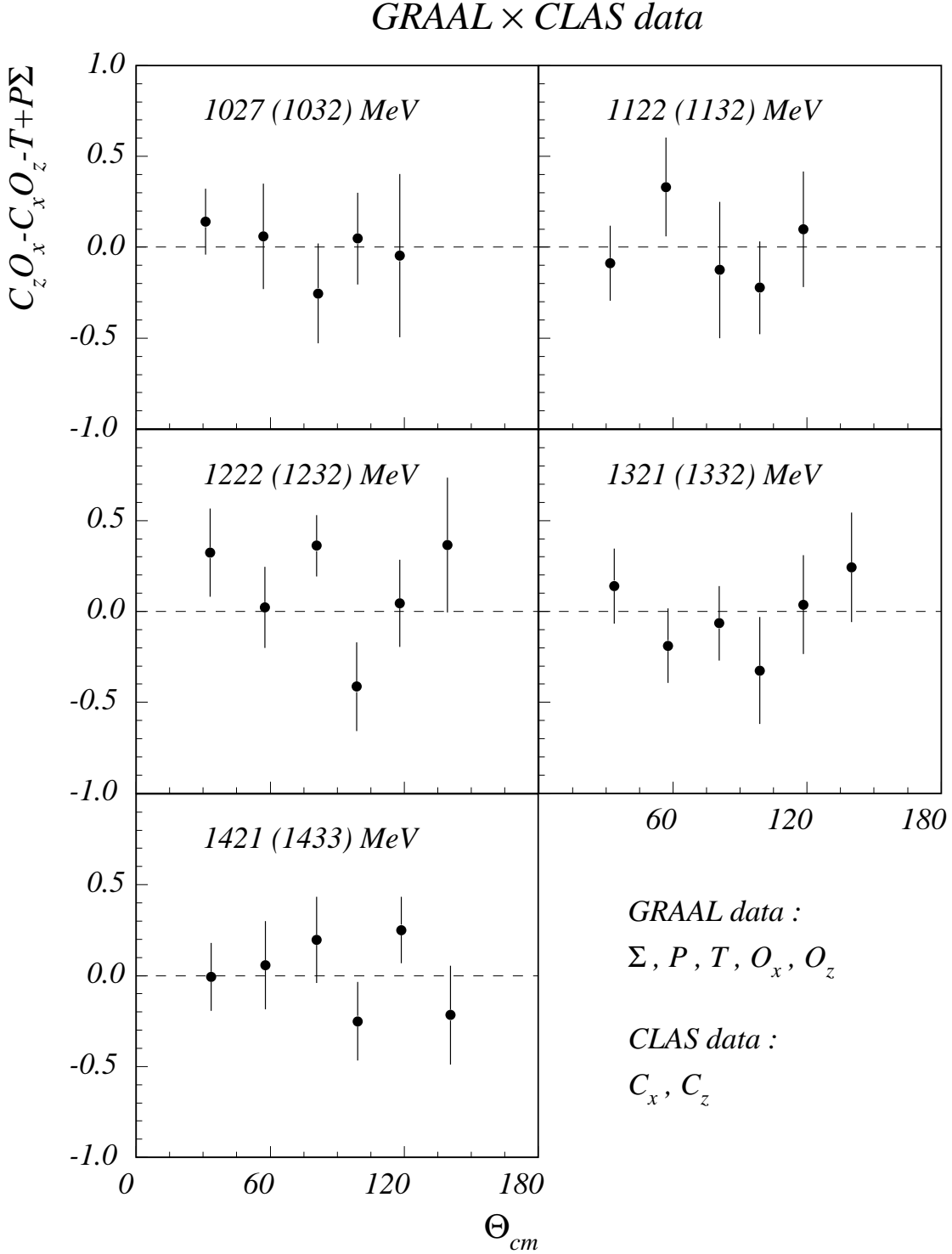} 
\end{center}
\caption{Angular distributions of the quantity $C_z O_x - C_x O_z - T + P \Sigma$. 
This quantity is calculated using the $C_x$ and $C_z$ results published by the CLAS collaboration (energy in parentheses) 
combined with our $O_x$ and $O_z$ data converted by eq. \ref{eqconv} to have the same $\hat{z}'$ axis convention 
and with our $\Sigma$, $P$ and $T$ measurements. The used 
CLAS data are those corresponding to the angles $\cos\theta_{cm}$=0.85, mean(0.65,0.45), mean(0.25,0.05), -0.15, 
mean(-0.35,-0.55) and -0.75. We should have the equality $C_z O_x - C_x O_z - T + P \Sigma = 0$ (eq. \ref{eqobs1b}).}
\label{pcokl}
\end{figure}

Since all observables entering in eqs. \ref{eqobs1} and \ref{eqobs1b} were measured either by GRAAL 
($\Sigma$, $P$, $T$, $O_x$, $O_z$) or by CLAS ($P$, $C_x$, $C_z$ - their $P$ data were confirmed by our
measurements \cite{lle07}), the
two sets of data can be therefore compared and combined. Within the error bars,
the agreement between the two sets of equal combinations $(1+T^2-\Sigma^2-O_x^2-O_z^2)^{1/2}$ (GRAAL) and 
$(P^2+C_x^2+C_z^2)^{1/2}$ (CLAS) is fair (fig. \ref{pckl}) and tends to confirm the previously observed
saturation to the value 1 of $R=(P^2+C_x^2+C_z^2)^{1/2}$, whatever the energy or angle.
Fig. \ref{pcokl} displays the values for the combined GRAAL-CLAS quantity $C_z O_x-C_x O_z-T+P\Sigma$.
Within the uncertainties, the expected value (1) is obtained, confirming again the overall consistency of
the GRAAL and CLAS data.

It has been demonstrated \cite{chi97} that the knowledge of the unpolarized
cross section, the three single-spin observables and at least four double-spin observables - provided
they have not all the same type - is sufficient to determine uniquely the four complex reaction amplitudes.
Therefore, only one additional double polarization observable measured using a polarized target
will suffice to extract unambiguously these amplitudes.

\subsection{Comparison to models}

\begin{figure}
\begin{center}
\includegraphics[width=0.9\linewidth]{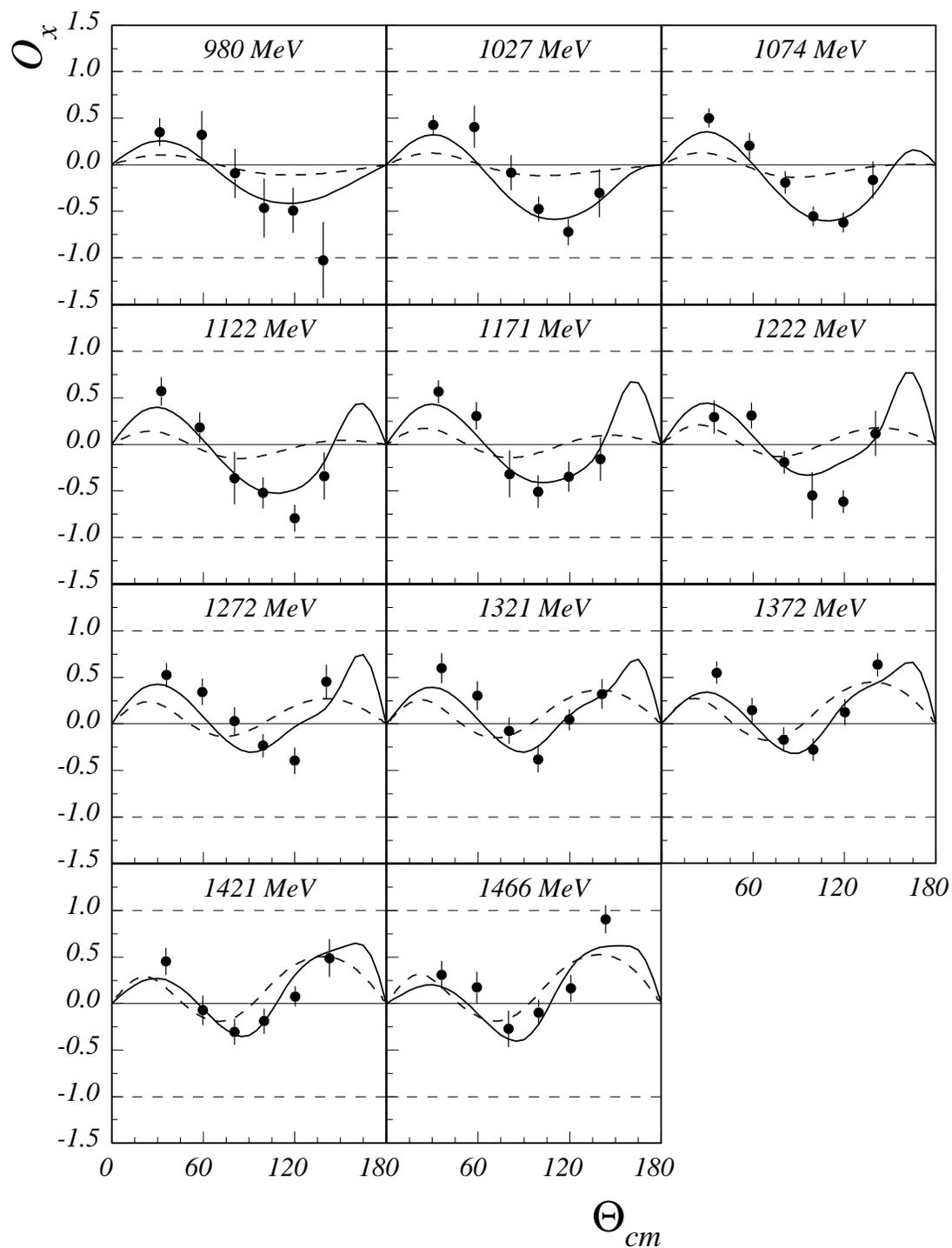} 
\end{center}
\caption{Angular distributions of the beam recoil observable $O_x$.
Data are compared with the predictions 
of the BG (solid line) and RPR (dashed line) models.}
\label{oxkl}
\end{figure}

\begin{figure}
\begin{center}
\includegraphics[width=0.9\linewidth]{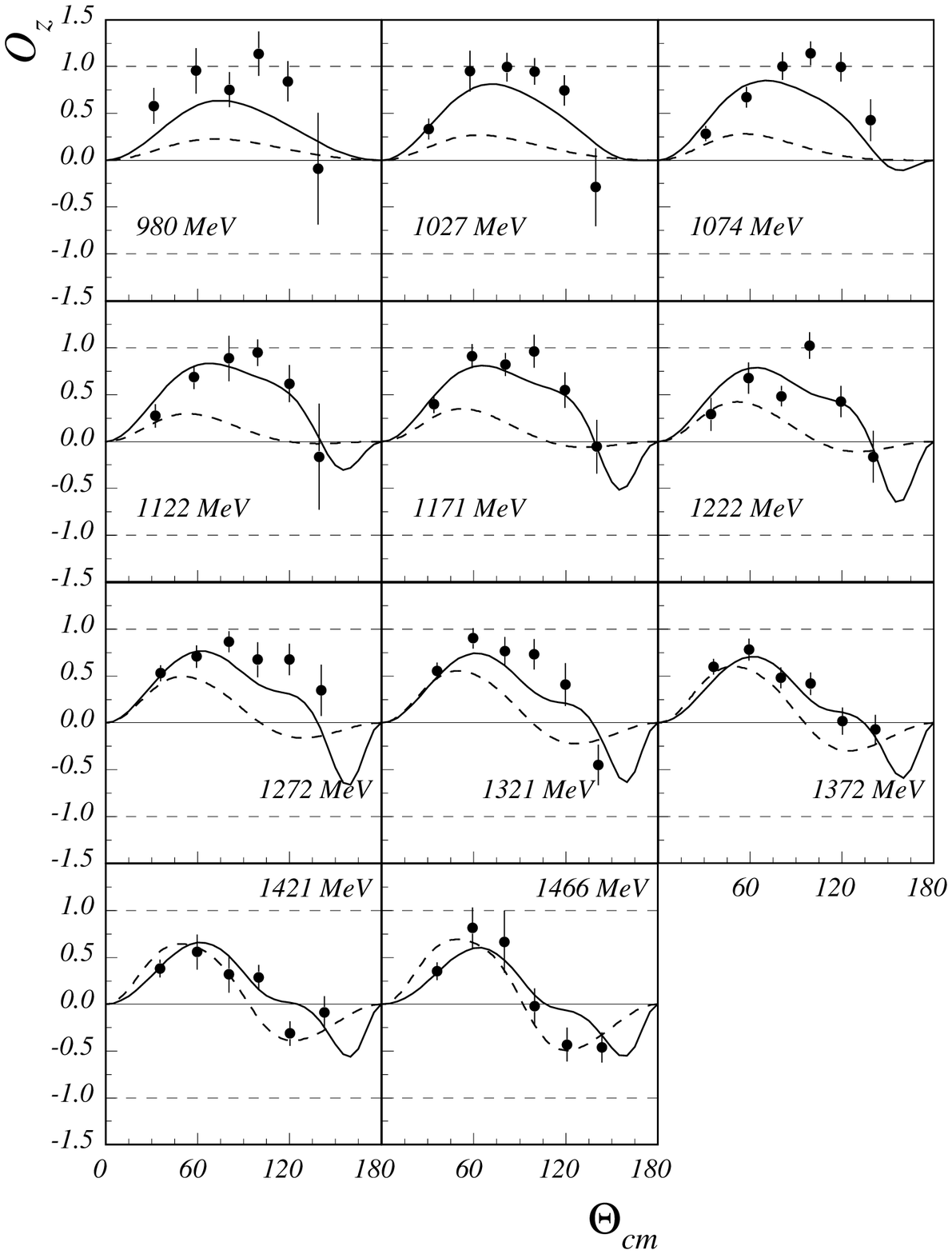} 
\end{center}
\caption{Angular distributions of the beam recoil observable $O_z$. 
Data are compared with the predictions 
of the BG (solid line) and RPR (dashed line) models.}
\label{ozkl}
\end{figure}

\begin{figure}
\begin{center}
\includegraphics[width=0.9\linewidth]{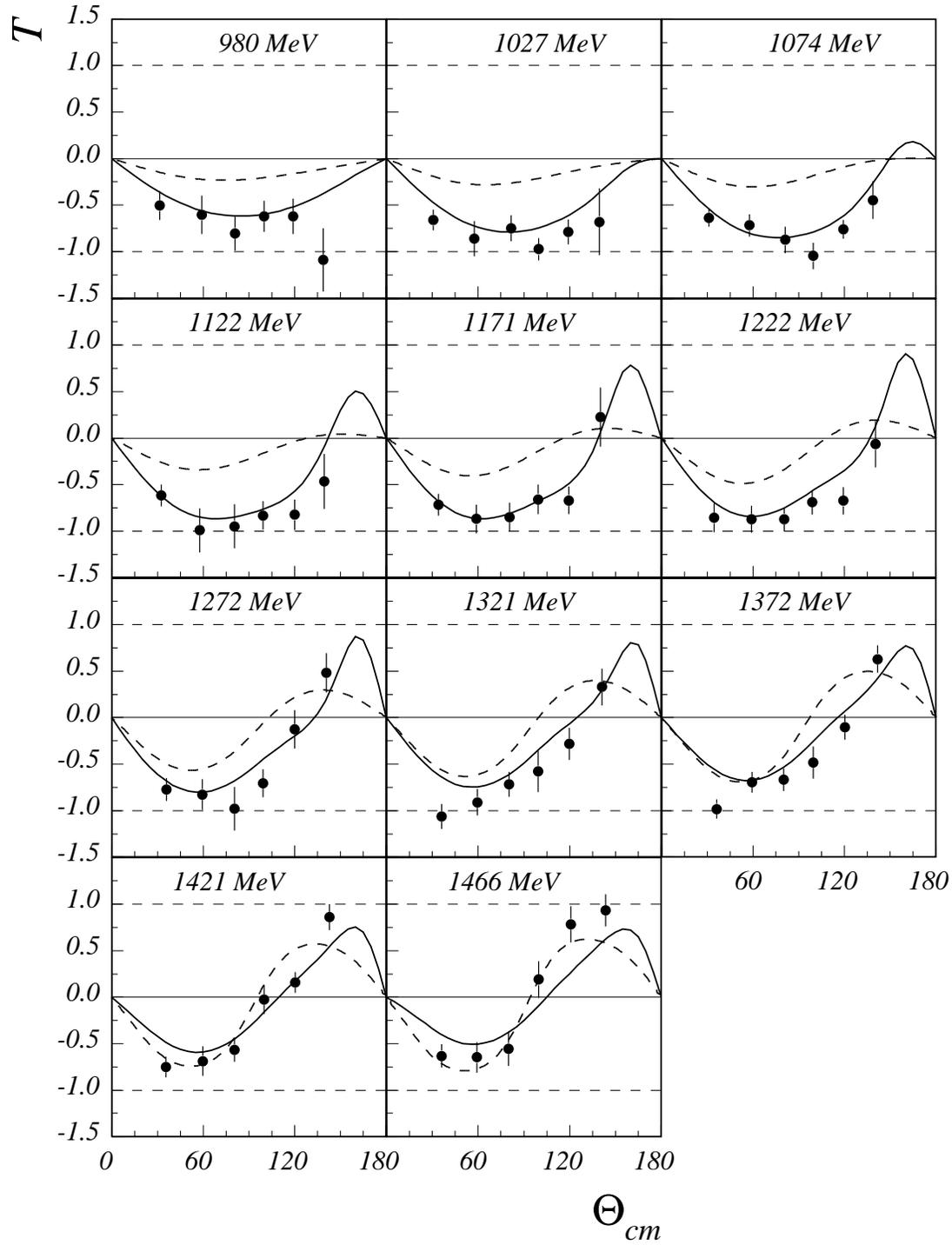} 
\end{center}
\caption{Angular distributions of the target asymmetry $T$.
Data are compared with the predictions 
of the BG (solid line) and RPR (dashed line) models.}
\label{tkl}
\end{figure}

We have compared our results with two models: the Ghent isobar RPR 
(Regge-plus-resonance) model \cite{cor06}-\cite{cov08} and the coupled-channel partial wave analysis developed by 
the Bonn-Gatchina collaboration \cite{ani05}-\cite{sar08}. In the following, these models will be refered as RPR 
and BG, respectively. The comparison is shown in figs. \ref{oxkl} to \ref{tkl}.

The RPR model is an isobar model for $K\Lambda$ photo- and electroproduction. 
In addition to the Born and kaonic contributions, it includes a Reggeized 
t-channel background which is fixed to high-energy data. 
The fitted database includes differential cross section, beam asymmetry and recoil polarization 
photoproduction results. The model variant presented here contains, besides the known $N^*$ resonances 
($S_{11}$(1650), $P_{11}$(1710), $P_{13}$(1720)), the $P_{13}$(1900) state (** in the PDG \cite{pdg04}) 
and a missing $D_{13}$(1900) resonance. This solution was found 
to provide the best overall agreement with the combined photo- and electroproduction database. 
As one can see in figs. \ref{oxkl} to \ref{tkl}, the RPR prediction (dashed line) qualitatively reproduces 
all observed structures. Interestingly enough, the model best reproduces the data at high energy (1400-1500 MeV), 
where the $P_{13}$(1900) and $D_{13}$(1900) contributions are maximal.

The BG model is a combined analysis of experiments with $\pi N$, $\eta N$, 
$K\Lambda$ and $K\Sigma$ final states. As compared to the other models,
this partial-wave analysis takes into account a much larger database which includes
most of the available results (differential cross sections and polarization observables).
For the $\gamma p \rightarrow K^+\Lambda$ reaction, the main resonant contributions come from the 
$S_{11}$(1535), $S_{11}$(1650), $P_{13}$(1720), $P_{13}$(1900) and $P_{11}$(1840) resonances.
To achieve a good description of the recent $C_x$ and $C_z$ CLAS measurements, the ** $P_{13}$(1900)
had to be introduced. It should be noted that, at this stage of the analysis, the contribution 
of the missing $D_{13}$(1900) is significantly reduced as compared to previous versions of the model.
As shown in figs. \ref{oxkl}-\ref{tkl}, this last version (solid line) provides a good overall agreement.
On the contrary, the solution without the $P_{13}$(1900) (not shown) fails to reproduce the data.

More refined analyses with the RPR and BG models are in progress and will be published later on.
Comparison with the dynamical coupled-channel model of Saclay-Argonne-Pittsburgh 
\cite{jul06}-\cite{sag08} has also started.

\section{Summary}

In this paper, we have presented new results for the reaction $\gamma p \rightarrow K^+\Lambda$
from threshold to $E_\gamma \sim$ 1500 MeV. Measurements of the beam-recoil observables $O_x$, $O_z$ and target
asymmetries $T$ were obtained over a wide angular range. 
We have compared our results with two isobar models which are in reasonable agreement with the whole
data set. They both confirm the necessity to introduce new or poorly known resonances in the  1900 MeV  mass region
($P_{13}$ and/or $D_{13}$).

It should be underlined that from now on only one additional double polarization 
observable (beam-target or target-recoil) would be sufficient to extract the four helicity 
amplitudes of the reaction.

\vspace{5mm}

\noindent
{\bf Acknowledgements}

\noindent
We are grateful to A.V. Sarantsev, B. Saghai, T. Corthals, J. Ryckebusch and P. Vancraeyveld
for communication of their most recent analyses and J.M. Richard for fruitful discussions on
the spin observable constraints. We thank R. Schumacher for communication of the CLAS
data. The support of the technical groups from all contributing
institutions is greatly acknowledged. It is a pleasure to thank the ESRF as a host institution
and its technical staff for the smooth operation of the storage ring.

\newpage

\onecolumn

\newpage

\newpage

\newpage

\newpage

\newpage

\newpage

\newpage

\onecolumn

\begin{table}
\begin{center}
\caption{Beam-recoil $O_x$ values.}
\label{table_oxkl}

\vspace{0.3cm}

\begin{tabular}{|r|r|r|r|r|r|r|r|} \hline

\multicolumn{1}{|c|}{$\theta_{cm} (^o)$} & \multicolumn{1}{|c|}{$E_{\gamma}$=980 MeV} &
\multicolumn{1}{|c|}{$\theta_{cm} (^o)$} & \multicolumn{1}{|c|}{$E_{\gamma}$=1027 MeV} &
\multicolumn{1}{|c|}{$\theta_{cm} (^o)$} & \multicolumn{1}{|c|}{$E_{\gamma}$=1074 MeV} &
\multicolumn{1}{|c|}{$\theta_{cm} (^o)$} & \multicolumn{1}{|c|}{$E_{\gamma}$=1122 MeV} \\ \hline

  31.3 &  0.349 $\pm$ 0.150 &   30.6 &  0.425 $\pm$ 0.108 &   31.2 &  0.502 $\pm$ 0.103 &   32.4 &  0.570 $\pm$ 0.154 \\
  59.1 &  0.320 $\pm$ 0.255 &   57.5 &  0.408 $\pm$ 0.225 &   57.5 &  0.202 $\pm$ 0.140 &   57.6 &  0.179 $\pm$ 0.161 \\
  80.7 & -0.094 $\pm$ 0.262 &   81.7 & -0.085 $\pm$ 0.189 &   81.0 & -0.190 $\pm$ 0.120 &   80.6 & -0.365 $\pm$ 0.282 \\
  99.8 & -0.464 $\pm$ 0.320 &   99.8 & -0.477 $\pm$ 0.133 &   99.7 & -0.552 $\pm$ 0.106 &   99.2 & -0.522 $\pm$ 0.165 \\
 118.9 & -0.490 $\pm$ 0.244 &  119.0 & -0.723 $\pm$ 0.142 &  119.3 & -0.621 $\pm$ 0.105 &  119.9 & -0.795 $\pm$ 0.146 \\
 138.6 & -1.028 $\pm$ 0.410 &  139.5 & -0.304 $\pm$ 0.259 &  138.8 & -0.163 $\pm$ 0.200 &  139.3 & -0.341 $\pm$ 0.252 \\ \hline
 
\multicolumn{1}{|c|}{$\theta_{cm} (^o)$} & \multicolumn{1}{|c|}{$E_{\gamma}$=1171 MeV} &
\multicolumn{1}{|c|}{$\theta_{cm} (^o)$} & \multicolumn{1}{|c|}{$E_{\gamma}$=1222 MeV} &
\multicolumn{1}{|c|}{$\theta_{cm} (^o)$} & \multicolumn{1}{|c|}{$E_{\gamma}$=1272 MeV} &
\multicolumn{1}{|c|}{$\theta_{cm} (^o)$} & \multicolumn{1}{|c|}{$E_{\gamma}$=1321 MeV} \\ \hline

  34.1 &  0.567 $\pm$ 0.122 &   34.6 &  0.294 $\pm$ 0.179 &   35.8 &  0.526 $\pm$ 0.130 &   36.0 &  0.599 $\pm$ 0.161 \\
  58.9 &  0.306 $\pm$ 0.148 &   58.9 &  0.310 $\pm$ 0.139 &   59.2 &  0.345 $\pm$ 0.140 &   59.4 &  0.304 $\pm$ 0.155 \\
  80.5 & -0.319 $\pm$ 0.254 &   80.4 & -0.193 $\pm$ 0.124 &   80.6 &  0.028 $\pm$ 0.155 &   80.3 & -0.073 $\pm$ 0.142 \\
  99.3 & -0.510 $\pm$ 0.175 &   98.9 & -0.548 $\pm$ 0.252 &   99.2 & -0.232 $\pm$ 0.125 &   99.2 & -0.383 $\pm$ 0.141 \\
 119.4 & -0.347 $\pm$ 0.162 &  119.1 & -0.615 $\pm$ 0.121 &  119.9 & -0.395 $\pm$ 0.143 &  119.9 &  0.046 $\pm$ 0.114 \\
 140.4 & -0.160 $\pm$ 0.234 &  140.3 &  0.116 $\pm$ 0.242 &  140.8 &  0.454 $\pm$ 0.187 &  141.3 &  0.323 $\pm$ 0.158 \\ \hline

\end{tabular}

\begin{tabular}{|r|r|r|r|r|r|} \hline

\multicolumn{1}{|c|}{$\theta_{cm} (^o)$} & \multicolumn{1}{|c|}{$E_{\gamma}$=1372 MeV} &
\multicolumn{1}{|c|}{$\theta_{cm} (^o)$} & \multicolumn{1}{|c|}{$E_{\gamma}$=1421 MeV} &
\multicolumn{1}{|c|}{$\theta_{cm} (^o)$} & \multicolumn{1}{|c|}{$E_{\gamma}$=1466 MeV} \\ \hline

  36.1 &  0.552 $\pm$ 0.119 &   35.7 &  0.455 $\pm$ 0.144 &   35.9 &  0.307 $\pm$ 0.150 \\
  59.5 &  0.150 $\pm$ 0.130 &   59.6 & -0.072 $\pm$ 0.160 &   59.3 &  0.172 $\pm$ 0.171 \\
  80.1 & -0.168 $\pm$ 0.131 &   80.3 & -0.303 $\pm$ 0.139 &   80.0 & -0.270 $\pm$ 0.195 \\
  99.4 & -0.276 $\pm$ 0.122 &   99.7 & -0.190 $\pm$ 0.137 &   99.7 & -0.096 $\pm$ 0.139 \\
 120.4 &  0.124 $\pm$ 0.138 &  120.4 &  0.076 $\pm$ 0.110 &  120.8 &  0.164 $\pm$ 0.145 \\
 141.9 &  0.636 $\pm$ 0.126 &  142.8 &  0.490 $\pm$ 0.205 &  143.7 &  0.905 $\pm$ 0.152 \\ \hline

\end{tabular}

\end{center}
\end{table}

\begin{table}
\begin{center}
\caption{Beam-recoil $O_z$ values.}
\label{table_ozkl}

\vspace{0.3cm}

\begin{tabular}{|r|r|r|r|r|r|r|r|} \hline

\multicolumn{1}{|c|}{$\theta_{cm} (^o)$} & \multicolumn{1}{|c|}{$E_{\gamma}$=980 MeV} &
\multicolumn{1}{|c|}{$\theta_{cm} (^o)$} & \multicolumn{1}{|c|}{$E_{\gamma}$=1027 MeV} &
\multicolumn{1}{|c|}{$\theta_{cm} (^o)$} & \multicolumn{1}{|c|}{$E_{\gamma}$=1074 MeV} &
\multicolumn{1}{|c|}{$\theta_{cm} (^o)$} & \multicolumn{1}{|c|}{$E_{\gamma}$=1122 MeV} \\ \hline

  31.3 &  0.581 $\pm$ 0.194 &   30.6 &  0.333 $\pm$ 0.110 &   31.2 &  0.285 $\pm$ 0.080 &   32.4 &  0.274 $\pm$ 0.124 \\
  59.1 &  0.956 $\pm$ 0.242 &   57.5 &  0.951 $\pm$ 0.216 &   57.5 &  0.674 $\pm$ 0.112 &   57.6 &  0.687 $\pm$ 0.127 \\
  80.7 &  0.754 $\pm$ 0.186 &   81.7 &  0.995 $\pm$ 0.154 &   81.0 &  1.003 $\pm$ 0.148 &   80.6 &  0.888 $\pm$ 0.244 \\
  99.8 &  1.139 $\pm$ 0.237 &   99.8 &  0.949 $\pm$ 0.140 &   99.7 &  1.140 $\pm$ 0.130 &   99.2 &  0.950 $\pm$ 0.144 \\
 118.9 &  0.841 $\pm$ 0.215 &  119.0 &  0.744 $\pm$ 0.162 &  119.3 &  0.996 $\pm$ 0.156 &  119.9 &  0.618 $\pm$ 0.197 \\
 138.6 & -0.091 $\pm$ 0.597 &  139.5 & -0.287 $\pm$ 0.415 &  138.8 &  0.427 $\pm$ 0.223 &  139.3 & -0.162 $\pm$ 0.568 \\ \hline

\multicolumn{1}{|c|}{$\theta_{cm} (^o)$} & \multicolumn{1}{|c|}{$E_{\gamma}$=1171 MeV} &
\multicolumn{1}{|c|}{$\theta_{cm} (^o)$} & \multicolumn{1}{|c|}{$E_{\gamma}$=1222 MeV} &
\multicolumn{1}{|c|}{$\theta_{cm} (^o)$} & \multicolumn{1}{|c|}{$E_{\gamma}$=1272 MeV} &
\multicolumn{1}{|c|}{$\theta_{cm} (^o)$} & \multicolumn{1}{|c|}{$E_{\gamma}$=1321 MeV} \\ \hline

  34.1 &  0.398 $\pm$ 0.093 &   34.6 &  0.291 $\pm$ 0.177 &   35.8 &  0.532 $\pm$ 0.087 &   36.0 &  0.554 $\pm$ 0.090 \\
  58.9 &  0.914 $\pm$ 0.128 &   58.9 &  0.678 $\pm$ 0.167 &   59.2 &  0.710 $\pm$ 0.119 &   59.4 &  0.904 $\pm$ 0.108 \\
  80.5 &  0.825 $\pm$ 0.123 &   80.4 &  0.485 $\pm$ 0.109 &   80.6 &  0.867 $\pm$ 0.112 &   80.3 &  0.767 $\pm$ 0.153 \\
  99.3 &  0.964 $\pm$ 0.175 &   98.9 &  1.025 $\pm$ 0.143 &   99.2 &  0.676 $\pm$ 0.188 &   99.2 &  0.734 $\pm$ 0.161 \\
 119.4 &  0.550 $\pm$ 0.190 &  119.1 &  0.426 $\pm$ 0.166 &  119.9 &  0.677 $\pm$ 0.166 &  119.9 &  0.409 $\pm$ 0.229 \\
 140.4 & -0.055 $\pm$ 0.286 &  140.3 & -0.162 $\pm$ 0.276 &  140.8 &  0.349 $\pm$ 0.272 &  141.3 & -0.448 $\pm$ 0.217 \\ \hline

\end{tabular}

\begin{tabular}{|r|r|r|r|r|r|} \hline

\multicolumn{1}{|c|}{$\theta_{cm} (^o)$} & \multicolumn{1}{|c|}{$E_{\gamma}$=1372 MeV} &
\multicolumn{1}{|c|}{$\theta_{cm} (^o)$} & \multicolumn{1}{|c|}{$E_{\gamma}$=1421 MeV} &
\multicolumn{1}{|c|}{$\theta_{cm} (^o)$} & \multicolumn{1}{|c|}{$E_{\gamma}$=1466 MeV} \\ \hline

  36.1 &  0.600 $\pm$ 0.084 &   35.7 &  0.384 $\pm$ 0.094 &   35.9 &  0.354 $\pm$ 0.095 \\
  59.4 &  0.784 $\pm$ 0.119 &   59.6 &  0.558 $\pm$ 0.185 &   59.3 &  0.814 $\pm$ 0.222 \\
  80.1 &  0.484 $\pm$ 0.112 &   80.3 &  0.322 $\pm$ 0.195 &   80.0 &  0.666 $\pm$ 0.332 \\
  99.4 &  0.419 $\pm$ 0.120 &   99.7 &  0.289 $\pm$ 0.134 &   99.7 & -0.023 $\pm$ 0.192 \\
 120.4 &  0.019 $\pm$ 0.145 &  120.4 & -0.313 $\pm$ 0.131 &  120.8 & -0.432 $\pm$ 0.180 \\
 141.9 & -0.072 $\pm$ 0.159 &  142.8 & -0.085 $\pm$ 0.172 &  143.7 & -0.461 $\pm$ 0.162 \\ \hline

\end{tabular}

\end{center}
\end{table}

\begin{table}
\begin{center}
\caption{Target asymmetry $T$ values.}
\label{table_tkl}

\vspace{0.3cm}

\begin{tabular}{|r|r|r|r|r|r|r|r|} \hline

\multicolumn{1}{|c|}{$\theta_{cm} (^o)$} & \multicolumn{1}{|c|}{$E_{\gamma}$=980 MeV} &
\multicolumn{1}{|c|}{$\theta_{cm} (^o)$} & \multicolumn{1}{|c|}{$E_{\gamma}$=1027 MeV} &
\multicolumn{1}{|c|}{$\theta_{cm} (^o)$} & \multicolumn{1}{|c|}{$E_{\gamma}$=1074 MeV} &
\multicolumn{1}{|c|}{$\theta_{cm} (^o)$} & \multicolumn{1}{|c|}{$E_{\gamma}$=1122 MeV} \\ \hline

  31.3 & -0.506 $\pm$ 0.156 &   30.6 & -0.663 $\pm$ 0.112 &   31.2 & -0.635 $\pm$ 0.096 &   32.4 & -0.615 $\pm$ 0.118 \\
  59.1 & -0.607 $\pm$ 0.206 &   57.5 & -0.860 $\pm$ 0.190 &   57.5 & -0.718 $\pm$ 0.120 &   57.6 & -0.991 $\pm$ 0.237 \\
  80.7 & -0.803 $\pm$ 0.185 &   81.7 & -0.749 $\pm$ 0.139 &   81.0 & -0.874 $\pm$ 0.141 &   80.6 & -0.949 $\pm$ 0.239 \\
  99.8 & -0.622 $\pm$ 0.166 &   99.8 & -0.974 $\pm$ 0.121 &   99.7 & -1.048 $\pm$ 0.140 &   99.2 & -0.833 $\pm$ 0.153 \\
 118.9 & -0.622 $\pm$ 0.187 &  119.0 & -0.789 $\pm$ 0.133 &  119.3 & -0.760 $\pm$ 0.102 &  119.9 & -0.825 $\pm$ 0.165 \\
 138.6 & -1.090 $\pm$ 0.341 &  139.5 & -0.681 $\pm$ 0.359 &  138.8 & -0.448 $\pm$ 0.203 &  139.3 & -0.465 $\pm$ 0.296 \\ \hline
 
\multicolumn{1}{|c|}{$\theta_{cm} (^o)$} & \multicolumn{1}{|c|}{$E_{\gamma}$=1171 MeV} &
\multicolumn{1}{|c|}{$\theta_{cm} (^o)$} & \multicolumn{1}{|c|}{$E_{\gamma}$=1222 MeV} &
\multicolumn{1}{|c|}{$\theta_{cm} (^o)$} & \multicolumn{1}{|c|}{$E_{\gamma}$=1272 MeV} &
\multicolumn{1}{|c|}{$\theta_{cm} (^o)$} & \multicolumn{1}{|c|}{$E_{\gamma}$=1321 MeV} \\ \hline

  34.1 & -0.715 $\pm$ 0.116 &   34.6 & -0.858 $\pm$ 0.155 &   35.8 & -0.773 $\pm$ 0.123 &   36.0 & -1.064 $\pm$ 0.133 \\
  58.9 & -0.869 $\pm$ 0.154 &   58.9 & -0.874 $\pm$ 0.145 &   59.2 & -0.827 $\pm$ 0.166 &   59.4 & -0.910 $\pm$ 0.142 \\
  80.5 & -0.850 $\pm$ 0.158 &   80.4 & -0.871 $\pm$ 0.124 &   80.6 & -0.979 $\pm$ 0.234 &   80.3 & -0.716 $\pm$ 0.133 \\
  99.3 & -0.659 $\pm$ 0.159 &   98.9 & -0.690 $\pm$ 0.132 &   99.2 & -0.707 $\pm$ 0.150 &   99.2 & -0.576 $\pm$ 0.223 \\
 119.4 & -0.669 $\pm$ 0.150 &  119.1 & -0.675 $\pm$ 0.145 &  119.9 & -0.125 $\pm$ 0.208 &  119.9 & -0.281 $\pm$ 0.174 \\
 140.4 &  0.226 $\pm$ 0.316 &  140.3 & -0.066 $\pm$ 0.249 &  140.8 &  0.482 $\pm$ 0.213 &  141.3 &  0.331 $\pm$ 0.198 \\ \hline
  
\end{tabular}

\begin{tabular}{|r|r|r|r|r|r|} \hline

\multicolumn{1}{|c|}{$\theta_{cm} (^o)$} & \multicolumn{1}{|c|}{$E_{\gamma}$=1372 MeV} &
\multicolumn{1}{|c|}{$\theta_{cm} (^o)$} & \multicolumn{1}{|c|}{$E_{\gamma}$=1421 MeV} &
\multicolumn{1}{|c|}{$\theta_{cm} (^o)$} & \multicolumn{1}{|c|}{$E_{\gamma}$=1466 MeV} \\ \hline

  36.1 & -0.983 $\pm$ 0.104 &   35.7 & -0.753 $\pm$ 0.112 &   35.9 & -0.632 $\pm$ 0.127 \\
  59.5 & -0.695 $\pm$ 0.113 &   59.6 & -0.687 $\pm$ 0.159 &   59.3 & -0.648 $\pm$ 0.166 \\
  80.1 & -0.669 $\pm$ 0.123 &   80.3 & -0.564 $\pm$ 0.131 &   80.0 & -0.553 $\pm$ 0.185 \\
  99.4 & -0.482 $\pm$ 0.175 &   99.7 & -0.025 $\pm$ 0.157 &   99.7 &  0.190 $\pm$ 0.196 \\
 120.4 & -0.104 $\pm$ 0.135 &  120.4 &  0.160 $\pm$ 0.112 &  120.8 &  0.785 $\pm$ 0.195 \\
 141.9 &  0.629 $\pm$ 0.147 &  142.8 &  0.859 $\pm$ 0.140 &  143.7 &  0.933 $\pm$ 0.175 \\ \hline
  
\end{tabular}

\end{center}
\end{table}


\begin{thebibliography}{27}

\bibitem{bra06} R.K.Bradford {\it et al.}, Phys. Rev. C {\bf 73}, 035202 (2006).

\bibitem{gla04} K.-H. Glander {\it et al.}, Eur. Phys. J. A {\bf 19}, 251 (2004).

\bibitem{nab04} J.W.C. McNabb {\it et al.}, Phys. Rev. C {\bf 69}, 042201(R) (2004).

\bibitem{sum06} M. Sumihama {\it et al.}, Phys. Rev. C {\bf 73}, 035214 (2006).

\bibitem{lle07} A. Lleres {\it et al.}, Eur. Phys. J. A {\bf 31}, 79 (2007).

\bibitem{zeg03} R.G.T. Zegers {\it et al.}, Phys. Rev. Lett. {\bf 91}, 092001 (2003).

\bibitem{bra07} R.K.Bradford {\it et al.}, Phys. Rev. C {\bf 75}, 035205 (2007).

\bibitem{bar05} O. Bartalini {\it et al.}, Eur. Phys. J. A {\bf 26}, 399 (2005).

\bibitem{pdg04} {\it Review of Particle Physics 2004}, Phys. Lett. B {\bf 592}, 1 (2004).

\bibitem{ade90} R.A. Adelseck and B. Saghai, Phys. Rev. C {\bf 42}, 108 (1990).

\bibitem{bar75} I.S. Barker, A. Donnachie and J.K. Storrow, Nucl. Phys. B {\bf 95}, 347 (1975).

\bibitem{cal97} P. Calvat, Thesis, Universit\'e J. Fourier Grenoble, 1997, unpublished.

\bibitem{lee57} T.D. Lee and C.N. Yang, Phys. Rev. {\bf 108}, 1645 (1957).

\bibitem{chi97} W.T. Chiang and F. Tabakin, Phys. Rev. C {\bf 55}, 2054 (1997).

\bibitem{art07} X. Artru, J.M. Richard and J. Soffer, Phys. Rev. C {\bf 75}, 024002 (2007).

\bibitem{cor06} T. Corthals, J. Ryckebusch and T. Van Cauteren, Phys. Rev. C {\bf 73}, 045207 (2006).

\bibitem{cor07} T. Corthals, Thesis, Universiteit Gent, 2007, unpublished.

\bibitem{cor08} T. Corthals {\it et al.}, Phys. Lett. B {\bf 656}, 186 (2007).

\bibitem{cov08} T. Corthals, J. Ryckebusch and P. Vancraeyveld, private communication.

\bibitem{ani05} A.V. Anisovich {\it et al.}, Eur. Phys. J. A. {\bf 25}, 427 (2005).

\bibitem{sar05} A.V. Sarantsev {\it et al.}, Eur. Phys. J. A. {\bf 25}, 441 (2005).

\bibitem{ani07} A.V. Anisovich {\it et al.},  Eur. Phys. J A {\bf 34}, 243 (2007).

\bibitem{nik07} V.A. Nikonov {\it et al.}, arXiv:hep-ph/0707.3600 (2007).

\bibitem{sar08} A. Sarantsev, private communication.

\bibitem{jul06} B. Juli\'a-D\'iaz, B. Saghai, T.-S.H. Lee and F. Tabakin, Phys. Rev. C {\bf 73}, 055204 (2006).

\bibitem{sag07} B. Saghai, J.C. David, B. Juli\'a-D\'iaz and T.-S.H. Lee, Eur. Phys. J. A {\bf 31}, 512 (2007).

\bibitem{sag08} B. Saghai, private communication.

\end{thebibliography}
\end{document}